\PassOptionsToPackage{margin=1in}{geometry}
\documentclass[12pt]{article}

\usepackage{amssymb,amsmath,amsfonts,eurosym,geometry,ulem,graphicx,caption,color,setspace,sectsty,comment,footmisc,caption,natbib,pdflscape,subfigure,array,hyperref}

\usepackage[utf8]{inputenc} % allow utf-8 input
\usepackage[T1]{fontenc}    % use 8-bit T1 fonts
\usepackage{hyperref}       % hyperlinks
\usepackage{url}            % simple URL typesetting
\usepackage{booktabs}       % professional-quality tables
\usepackage{amsfonts}       % blackboard math symbols
\usepackage{nicefrac}       % compact symbols for 1/2, etc.
\usepackage{microtype}      % microtypography
\usepackage{lipsum}		% Can be removed after putting your text content
\usepackage{graphicx}
\usepackage{natbib}
\usepackage{doi}
\usepackage{amsmath}
\usepackage[margin=1in]{geometry}
\usepackage{siunitx}
\usepackage{caption}
\usepackage{multirow}
\usepackage{adjustbox}
\usepackage{longtable}
\usepackage{rotating}
\usepackage{tabularx}

\newcolumntype{L}[1]{>{\raggedright\let\newline\\arraybackslash\hspace{0pt}}m{#1}}
\newcolumntype{C}[1]{>{\centering\let\newline\\arraybackslash\hspace{0pt}}m{#1}}
\newcolumntype{R}[1]{>{\raggedleft\let\newline\\arraybackslash\hspace{0pt}}m{#1}}

\begin{document}

\begin{titlepage}
\title{Bellwether Trades: Characteristics of Trades influential in
Predicting Future Price Movements in Markets}
\author{Tejas Ramdas\thanks{Department of Statistics \& Data Science and Johnson Graduate School of Management, Cornell University. email for correspondence: tr336@cornell.edu} \and Martin T. Wells\thanks{Department of Statistics \& Data Science, Cornell University}}
\date{\today}
\maketitle
\begin{abstract}
\noindent In this study, we leverage powerful non-linear machine learning methods to identify the characteristics of trades that contain valuable information. First, we demonstrate the effectiveness of our optimized neural network predictor in accurately predicting future market movements. Then, we utilize the information from this successful neural network predictor to pinpoint the individual trades within each data point (trading window) that had the most impact on the optimized neural network's prediction of future price movements. This approach helps us uncover important insights about the heterogeneity in information content provided by trades of different sizes, venues, trading contexts, and over time.\\
\vspace{0in}
\vspace{0in}

\bigskip
\end{abstract}
\setcounter{page}{0}
\thispagestyle{empty}
\end{titlepage}
\pagebreak \newpage

\doublespacing

\section{Introduction} \label{sec:introduction}
The study of market frictions, variables linked to understanding market frictions and the study of the information content of certain types of trades have been important areas of focus in market microstructure research. A large body of literature, starting at least as far back as the seminal models of \cite{e2104c33-2a9d-3e4e-be9a-35cae4feb70d}, \cite{GLOSTEN198571} studies how trades with varying attributes may contribute differently to future market movements. Further, the information content offered by trades with specific attributes has been shown to be of consequence to the study of optimal actions of market centers, for example,  \cite{https://doi.org/10.1111/jofi.12185}). \cite{RePEc:eee:jfinec:v:34:y:1993:i:3:p:281-305}, \cite{CHAKRAVARTY2001289}, \cite{article}, among other studies analyze implications resulting from heterogeneity in information content offered by trades of varying attributes. Considerable work, going at least as far back as \cite{https://doi.org/10.1111/j.1540-6261.1991.tb03749.x}, is devoted specifically to the empirical estimation of the persistent price impact of trades of varying attributes. Further studies (for example,  \cite{2f4daf45-d507-3e93-8916-77093703379f},  \cite{RePEc:eee:dyncon:v:36:y:2012:i:4:p:501-522},\cite{RePEc:taf:eurjfi:v:18:y:2012:i:9:p:841-864}, \cite{RePEc:eee:finmar:v:7:y:2004:i:1:p:1-25}) and more recently \cite{campigli2023measuring} study the information content of trades/trading process under various settings, using varying empirical frameworks. \\

In this study, we present a framework using interpretable methods in machine learning to identify characteristics of trades that most influence our optimized neural network's prediction of future price movements. We begin by demonstrating the efficacy of non-linear, data-intensive neural network models in accurately predicting future market movements. By incorporating approaches from the Automated Machine Learning (AutoML) literature, we optimize both the predictor's architecture and hyper-parameters in search for the best-in-class neural network predictor for our application. Subsequently, we introduce an interpretation framework designed to pinpoint the specific trades within each "trading-window" that exert the most influence on the neural network's predictions of price movements in the subsequent window (of a definite number of trades). For instance, in trading window 10, trades 1, 2, and 7 might be influential in driving the neural network's forecast of a positive price movement in the subsequent window of trades. Conversely, in trading window 17, trades 5, 6, and 8 could be crucial in shaping a positive or negative prediction. By extracting and analyzing the characteristics of these influential trades—such as trades 1, 2, and 7 in trading window 10, and trades 5, 6, and 8 in trading window 17 — we can uncover key stylized facts about the information content and influence of different trade types.  \\

Our proposed framework makes contributions in two dimensions. First, we add to the literature related to estimating the information content of trades. Similar to \cite{https://doi.org/10.1111/j.1540-6261.1991.tb03749.x} and related approaches, our framework to identify characteristics of trades that offer most predictability in identifying future market movements does not require assumptions typically associated with formal economic models. Our framework inherently also considers information from the trading process in general as opposed to individual trades, in parallel with the logic of \cite{2f4daf45-d507-3e93-8916-77093703379f}. Our proposed framework can be modified to consider different time-frames of analysis, by shortening or lengthening window periods of consideration. By comparing influence scores of trades under different settings for window sizes, our framework allows a comparison of the short term vs long term influence of trades. The quality of our estimates of the predictability offered by trades of varying attributes is quantifiable, and can be evaluated by examining the accuracy of the optimized predictors built in the first stage. The number of attributes (for example size, venue, time of trade and more) can be adjusted based on the research needs of the user. In our framework, two trades with the exact same attributes are allowed to have different influence scores, depending on the context in which the trades occur. The absence of formal modelling and the resulting flexibility means that estimates extracted from our approach have the potential to motivate theoretical questions in the domain. \\

Second, our work attempts to assimilate an optimization step to ensure robustness in analyses that require the use of prediction models in downstream analysis. Specifically, our approach integrates an additional step to select optimized candidate predictors to interpret, from an initially specified class of predictors. A growing number of studies in economics, finance, and management (and other fields) use predictions from machine learning models and feature importance scores that drive these predictions for downstream analyses (\cite{RePEc:oup:rfinst:v:34:y:2021:i:7:p:3316-3363.} in the microstructure literature, \cite{athey2019machine} review some prominent methods). In settings where predictions and information relating to features of successful predictors are used in downstream analysis, additional measurement error is induced on top of pre-existing irreducible errors when the predictor in question is not the optimal predictor in its class. The severity of this induced error typically depends on a range of factors, including the size of the class of predictors  \cite{Varian2014Big}, variability in performance over the elements of this class (\cite{Hastie2009Elements}), topology of decision boundaries (\cite{Goodfellow-et-al-2016}), and more. Existing studies evaluate the predictive performance of a number of models, and select the best performing candidate from this (naturally) limited set for downstream analyses, creating vulnerabilities that stem from unverifiable assumptions about the magnitude of irreducible error  (\cite{Bishop2006Pattern}). Our work demonstrates an attempt to ameliorate this problem by (1) incorporating an optimization step between our prediction and inference frameworks, and (2) by ensuring that this optimization is effective when search spaces are large and sometimes continuous. \\

Finally, our framework continues an increasingly prominent line of research that develops and leverages machine learning methods to address the challenges posed by a rapidly evolving market landscape. For example, \cite{easley2019microstructure} take a machine learning approach to demonstrate the importance of microstructure variables in understanding market frictions in a world dominated by high-frequency computerized trading. Recent work by  \cite{chinco2018}, \cite{rossi2018},  \cite{krauss2017}, \cite{gu2020},  \cite{deprado2018} further underscore the effectiveness of modern methods in machine learning for understanding the importance of microstructure variables, and in predicting future market movements. Studies by  \cite{heaton2017}, \cite{sirignano2019}, \cite{bianchi2020}, \cite{feng2019},  \cite{chen2020} have increasingly employed neural networks to address questions in asset pricing and market microstructure.  \cite{kelly2023} expand on this emerging body of literature in greater detail.   \\

Our approach generates several new stylized facts and reaffirms other important discoveries in the literature. We find that the distribution of how much predictability each trade offers about future market movements has a long right tail. A minority of "Bell-Weather" trades make large contributions towards our successful neural network's prediction of future market movements. We find that controlling for time of trade, year, trade price, venue, trade sizes, high/low volume environment and interactions, that ETF trades offer less information in predicting their own future price movements than do individual stocks. We further find that controlling for our other attributes, odd-lot trades offer very slightly more information \footnote{ No inference is to be made relating to the direction of future market movements.} on future market movements compared to round lot trades. We find that small, odd-lot trades may be significantly more influential in predicting future price movements in windows where future positive price movements were predicted than in windows where future negative price movements were predicted. The effect for trades of size between 101 and 999 shares is small, significant and negative as compared with round lot trades. Controlling for our other attributes, we find that trades of larger sizes offer substantially less information on future market movements compared with round lot trades, and that this effect is large and significant, shedding more light on the contrasting results of O'Hara, Yao \& Ye, (2012); Choe \& Hansch (2005), Barclay \& Warner (1993) and Chakravarty (2001). In rhyme with \cite{RePEc:oup:rfinst:v:14:y:2001:i:4:p:1153-81} \footnote{Gratitude to Gideon Saar for feedback that highlighted this connection.}, we find that trades conducted in windows where future positive market movements were predicted showed a higher influence score than trades conducted in windows where future negative market movements were predicted, controlling for our other attributes. We further find that these estimates vary significantly across years, with differences particularly notable in years where market action was more volatile. This observation lends credence to our submission that trades with the same attributes might vary considerably in the predictability they offer depending on the context in which the trades occur. In addition to these and other results, we offer estimates related to how trades conducted at certain venues (because of the selection of trades into venues resulting from heterogeneity in cost structure, varied selections of order types, membership and access regulations, technological infrastructure, cross platform integration and other reasons; analysis in likeness to \cite{https://doi.org/10.1111/j.1540-6261.1995.tb04054.x}) may be more (or less) influential in predicting future price movements than trades of the same attributes conducted at other venues. We discuss these and other results in detail in section 3.\\

The remainder of this paper is organized as follows. Section 2 describes the data, prediction methodology, and our approach to optimize the architecture and hyper-parameters of our predictor. This section also describes our approach to detecting influential trades from this optimized neural network predictor in each trading window. Later, we detail a regression-based framework to extract the characteristics of these influential trades. In Section 3, we present and discuss our results. In Section 4, we explore future research directions and provide concluding comments.

\section{Data} \label{sec:data}
\subsection{Initial Overview}

To determine the characteristics of trades that most influence predictions of future market movements, we develop a prediction-interpretation framework that consists of the following steps. First, we extract trades and quotes information from the NYSE Daily Trades and Quotes database. Then, we pre-process our raw data and train a neural network classifier using this constructed input set of trades. The architecture and hyper-parameters of our predictive neural network model are optimized to predict market future market movements in a manner that utilizes most predictive power the data offers. Then, combining this approach with methods from the Interpretable Machine Learning literature, we query our optimized neural network predictor to extract influence scores for each trade that forms a part of the input to the previously trained optimized neural network classifier. Finally, we describe a regression framework to make sense of the influence scores derived from the optimized neural network predictor (and each focal input trade from our large collection), to return results that indicate the characteristics of trades that most influence the predictions of the optimized neural classifier in predicting future market movements. 

\subsection{Raw data and Inputs for Prediction}

We use data from the New York Stock Exchange (NYSE) Daily Trades and Quotes (DTAQ) database for our analysis, covering five years from 2017 through 2021. For each stock-day within this time-frame, we initially extract all trades that occurred during market hours. Our sample contains a total of approximately 33 billion trades. We gather data on several aspects of each trade: the trade price, volume, venue where the trade was conducted, prevailing best bid and offer prices at that venue immediately before the trade, corresponding bid-offer sizes, the National Best Bid and Offer (NBBO) prices and sizes prevailing just before the trade was conducted, the trade's timestamp, the time difference in nanoseconds between the current and previous trade, and whether the security is a single company stock or an Exchange Traded Fund (ETF). We exclude all securities that are neither single company stocks nor Exchange Traded Funds. After extracting all candidate trades and their attributes for each stock-day, we organize the extracted trades into windows. The first part of our analysis aims to utilize the information from trades within a given window to predict price action in the subsequent window. We assess whether the stock has been traded at least 4,000 times that day for each day and stock. We include a stock-day in our analysis if the stock has been traded a minimum of 4,000 times (based on the number of trades, not volume). This step ensures that only securities capable of accommodating our largest trial window size (maximum window size of 1936 trades) are included in our analysis. A more detailed description of the inputs and output variables of the neural network predictor is provided in the next subsection. 

\subsubsection{Input Variables (Prediction)}

We organize the extracted set of trades into windows, each containing sets of trades. For this study, we experimented with multiple windows to see which window size offered the highest predictability, from a minimum of 289 trades to a maximum of 1936 trades. In practice, the characteristics of trades that influence the prediction of future market movements might be different when different window sizes are considered, and they might perhaps be useful material for future studies. We also note that a window size consisting of $n$ trades might imply different time periods in question for highly traded vs lightly traded securities. We include measures of liquidity in both our prediction models and regression framework to attempt to account for this difference. The number of trades per window is always chosen to be a perfect square so that the trades can be arranged into a square matrix to feed into a Convolutional Neural Network (CNN). The windows of input trades are non-overlapping, and no two windows contain the same trade. We ensure that each window contains trades in sequential time order, and that no window contains trades spanning two different trading days. For each trade, we extracted a total of 35 attributes. The attributes of interest are listed below. Please note that many of the values of the associated fields below are transformed to make the inputs more tractable to serve as inputs to a CNN \footnote{In sections 2.3 and 2.4, we discuss reasons that suggest CNNs may be particularly well suited for our analysis, and we discuss our approach to architecture selection and hyper-parameter optimization.}.

\begin{table}[ht!]
\centering
\begin{tabularx}{\linewidth}{lX}
\toprule
\textbf{Variable}   & \textbf{Description} \\
\midrule
Time                & Time when the trade was executed in nanosecond resolution \\
Venue               & 21 dummies for the venue the trade was transacted at \\
Bid\_Price          & Price of the current highest bid just prior to trade \\
Bid\_Size           & Size of the current highest bid just prior to trade \\
Offer\_Price        & Price of the current lowest offer just prior to trade \\
Offer\_Size         & Size of the current lowest offer just prior to trade \\
Trade\_Volume       & Volume of the trade \\
Trade\_Price        & Price at which the trade was executed \\
nbid\_Price         & National Best Bid Price just prior to trade \\
nbid\_Size          & National Best Bid Size just prior to trade \\
noffer\_Price       & National Best Offer Price just prior to trade \\
noffer\_Size        & National Best Offer Size just prior to trade \\
ETF                 & Indicator if the security is an ETF \\
year                & Year when the trade was executed \\
time\_diff          & Time difference from the previous trade in nanoseconds \\
\bottomrule
\end{tabularx}
\caption{Description of Trade Attributes}
\label{tab:variables}
\end{table}

The above listed attributes are arranged into a tensor for each trading window. The dimensions of each tensor trading-window depends on the number of trades in that window. Each input window-tensor always takes the form $m \times m \times 35$, where 35 is the number of "channels", where each channel corresponds to one of our above mentioned input variables. The $ijm$'th entry in each input window tensor is the $m$'th attribute of the trade in the $ij$'th position. The trades are arranged sequentially in a grid, so that the $m$ attributes of the first trade is take the position $1,1 \times m$ and the $m$ attributes of the final $n$'th trade take the position $\sqrt{n}, \sqrt{n} \times m$. An Illustration is shown in Figure 1. \\ 

\begin{figure}[ht!]
    \centering
    \includegraphics[width=8cm, height=8cm]{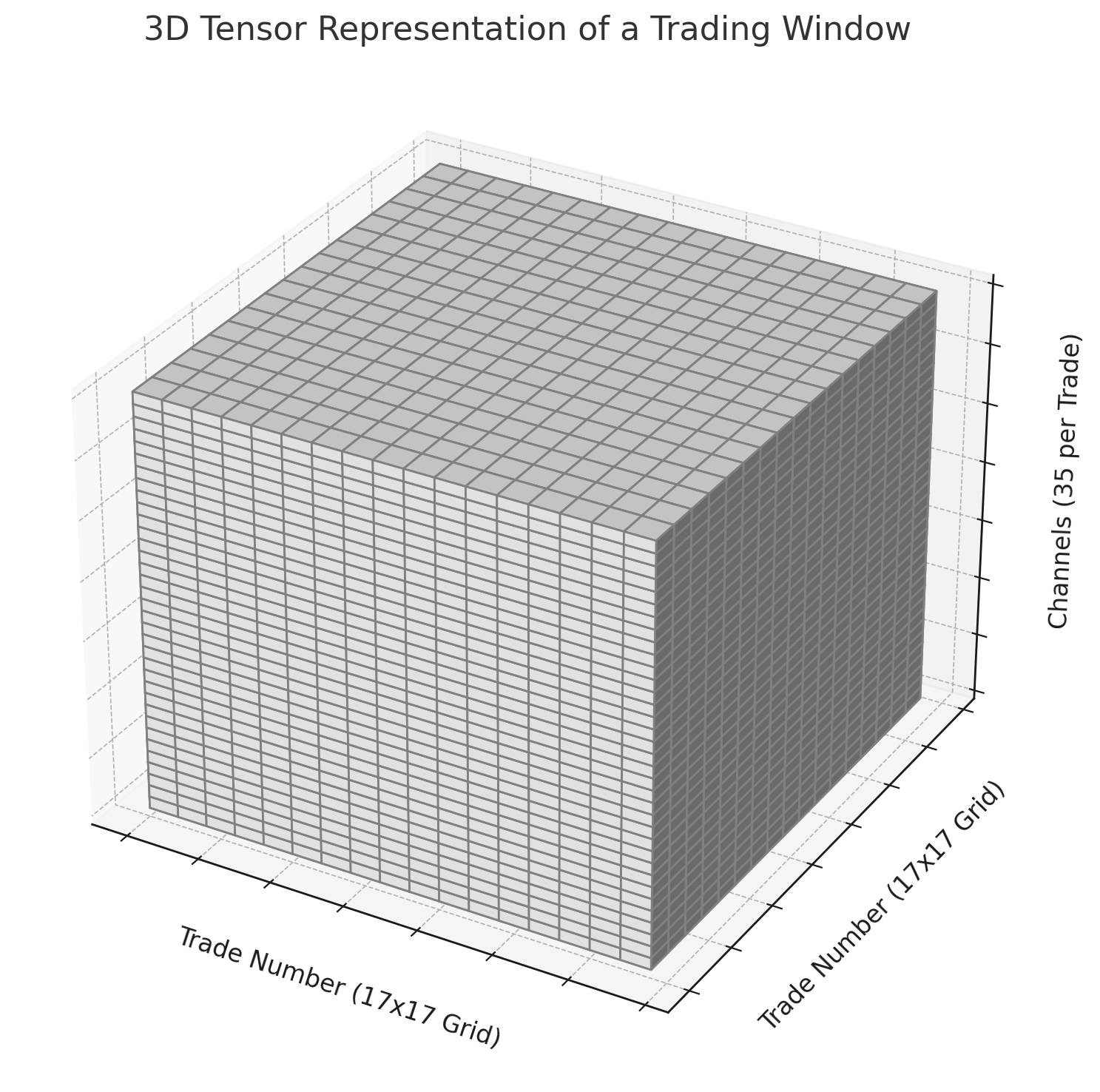}
    \caption{Each window has $n^2$ trades, arranged in a grid. Each of the 35 inputs of each trade are then $n \times n$ matrices arranged in a vector of dimensions $n \times n \times 35$}
    \label{fig:fig1}
\end{figure}

At this stage, it is important to note that we initially train optimal neural network models for various window sizes. We later present the best prediction accuracies obtained for each window size. We select the window size that offers the highest out-of-sample predictive power (289 trades per window) to extract the characteristics of influential trades. As an alternative to using sets of trades in a window, time and volume bars might also be helpful candidates for future study. However, using time or volume bars will preclude analysis and identification of the characteristics of consequential trades. Using time or volume bars shifts the discussion toward finding the characteristics of influential time or volume intervals rather than individual trades. 

\subsubsection{Outcome Variable (Prediction)}

For each trading window, we create an indicator variable that is assigned a value of $1$ if the average consolidated trade price of the subsequent window is greater than the trade price of the final trade of the previous window and $0$ otherwise. This is our variable of interest for \textit{prediction purposes}. Whereas we train a Neural Network to achieve high prediction accuracy on this variable of interest, our final objective is to use the information contained in this successful Neural Network to identify characteristics of trades (inputs) that are most influential in determining the predictions of the neural network. The design of the above indicator means that the last trading window on each trading day must be deleted from our analysis. Alternate outcome variables that capture different aspects of future market movements can potentially usefully supplement analysis conducted using our present outcome variable.  

\subsection{The Neural Network Architecture}

The objective for this subsection is to find a function $f(\cdot; \theta)$, such that $\hat{y}_{t+1} = f(\mathbf{x}^{(t)}; \mathbf{\theta})$ predicts future price movement $y_{t+1}$ accurately given input $\mathbf{x}^{(t)}$. $y_{t+1}$ is our response indicator. $y_{t+1}$, our response indicator, is 1 if the consolidated price of the subsequent trading window is higher than the final trade of the focal window, and 0 otherwise. We use an optimized CNN based architecture (\cite{6795724},  \cite{oshea2015introduction}, \cite{Goodfellow-et-al-2016}), for the following reasons. First, CNN architectures generally utilize fewer parameters than fully connected networks to treat inputs of similar dimensions, due to their architectural design (\cite{Goodfellow-et-al-2016}; \cite{zeiler2013visualizing}). Second, Recurrent Neural Network (RNN) architectures may have an inherent advantage over CNNs when inputs to the model are time dependent data, but this feature means that there is an inherent sequential dependency that constrains the ability of that predictor to compute outputs in parallel (\cite{Goodfellow-et-al-2016};   \cite{sze2017efficient}). Third, in contrast to newer, possibly more powerful architectures, an exhaustive and increasing body of literature has been dedicated towards interpretability and explainability in CNN architectures (\cite{DBLP:journals/corr/abs-2004-02806} review a part of this literature). Finally, the architecture of CNN is particularly well suited for, and has been widely successful in detecting patterns inside a grid-like topology (\cite{Goodfellow-et-al-2016}). However, using CNNs for our purpose comes with atleast one salient drawback: arranging trades in each input trade window to conform to the required grid-like topology means that artificial functional relationships may be created between unrelated trades. These artificial relationships generate some systemic noise both in the prediction and inference stages. We rely on our elephantine sample size to obviate this noise in our final estimates. Below, we describe basic elements of CNN architectures. Readers familiar with CNNs may safely skip to the next section. Consider an input tensor $\mathbf{X}^{(t)}$ of size $m \times m \times 35$ for each training window $t$. CNN architectures typically contain the following operations at their core.

1. \textbf{Convolutional Layers:} Let $f_{ij}^k(\mathbf{X}; \mathbf{W}^{(k)}, b^{(k)})$ represent the output of the $k$-th convolutional layer at position $(i, j)$, where $\mathbf{W}^{(k)}$ and $b^{(k)}$ are the weights and bias of the $k$-th layer. The convolution operation is defined as:
\[ f_{ij}^k(\mathbf{X}; \mathbf{W}^{(k)}, b^{(k)}) = \sigma\left(\sum_{p=1}^{P}\sum_{q=1}^{Q}\sum_{c=1}^{35}\mathbf{W}_{pqc}^{(k)} \cdot \mathbf{X}_{i+p, j+q, c} + b^{(k)}\right) \]
where $P \times Q$ is the size of the filter in the $k$-th layer, and $\sigma$ is a nonlinear activation function (for example ReLU). 

2. \textbf{Pooling Layers:} After each convolutional layer, a pooling layer is typically applied to reduce the spatial dimensions of the output. We employ the traditionally used "max pooling" approach of Zhou and Chellappa, (1988). Pooling makes the inputs robust to small translations of the input, so that changing the order of trades slightly and locally does not overly affect our predictions (Goodfellow, Bengio \& Courville, 2016).  For a pooling operation over a $2 \times 2$ window, the operation is defined as:
\[ g_{ij}^k = \text{pool}\left( f_{ij}^k, f_{i+1,j}^k, f_{i,j+1}^k, f_{i+1,j+1}^k \right) \]
where $\text{pool}$ is the max pooling function.

3. \textbf{Fully Connected Layers:} After a defined number of convolutional and pooling layers, the output is flattened and passed through one or more fully connected layers. Let $h^l$ denote the output of the $l$-th fully connected layer, defined as:
\[ h^l = \sigma\left( \mathbf{W}^{(l)} \cdot h^{l-1} + b^{(l)} \right) \]
with $h^0$ being the flattened output of the last pooling layer.

4. \textbf{Output Layer:} The final layer is typically a fully connected layer with a sigmoid activation function, producing the prediction probability $\hat{y}_{t+1}$ for the binary classification:
\[ \hat{y}_{t+1} = \text{sigmoid}\left( \mathbf{W}^{(L)} \cdot h^{L-1} + b^{(L)} \right) \]

We use the binary cross-entropy loss to train our classifier, defined as:
\[ \mathbf{\theta} = \underset{\mathbf{\theta}}{\mathrm{argmin}} \; - \frac{1}{n}\sum_{t=1}^{n} \left( y_{t+1} \log(\hat{y}_{t+1}) + (1 - y_{t+1}) \log(1 - \hat{y}_{t+1}) \right) \]
where $y_{t+1}$ is the ground truth label for window $t$, and $n$ is the number of training windows. The number and configuration of the different kinds of layers, and the hyper-parameters to describe each step of the CNN architecture is a crucial determinant of the network's final predictive power. We use a Bayesian approach from the Automated Machine Learning Literature, the Tree-Structured Parzen algorithm (\cite{article}), to optimize the architecture of the CNNs used to predict future price action. This optimization step, which we substantiate below, is a critical component of our methodological framework, as it must be satisfied that we have a performance optimized neural network to derive influence scores from (and that we do not query influence scores from simply any arbitrarily chosen neural network).  

\subsection{Hyper-parameter Optimization}

In this section, we discuss our approach to selecting an optimal architecture from the class of CNN architectures. The Tree-Structured Parzen Estimator (TPE) algorithm is a widely prevalent choice for architecture optimization in the Auto-ML literature (\cite{NIPS2011_86e8f7ab, watanabe2023treestructured}). The TPE algorithm supports optimization in cases where inputs may contain both continuous and discrete elements (\cite{NIPS2011_86e8f7ab}), has been found to approach optimal hyper-parameter values much faster than random grid searches and a wide class of other approaches (\cite{watanabe2023treestructured}), and is supported by infrastructure that significantly aids implementation (\cite{akiba2019optuna}). Given a class of predictors, in our case the class of convolutional neural network architectures $f(\cdot; \theta)$, we want to find a $\theta^*$ that maximizes the validation accuracy of the predictor. Let \( L(\theta) \) denote the objective function, in our case the validation accuracy, where \( \theta \) represents the set of hyper-parameters under evaluation. The Tree-Structured Parzen Estimator (TPE) is a model-based optimization method that constructs probabilistic models of the objective function to guide the selection of hyper-parameters. Specifically, the TPE models two probability distributions conditional on the performance of evaluated hyper-parameters
\[
l(\theta) = P(\theta \mid L(\theta) < L^*)
\]
which represents the probability distribution of hyper-parameters that result in objective function values less than a threshold \( L^* \) and the term
\[
g(\theta) = P(\theta \mid L(\theta) \geq L^*)
\]
represents the probability distribution of hyper-parameters that result in objective function values greater than or equal to \( L^* \). A typical choice for the threshold \( L^* \) is the median of the observed validation accuracies \( L(\theta) \) from previous evaluations. The next set of hyper-parameters to evaluate are chosen by maximizing the Expected Improvement (EI), a criterion that quantifies the expected reduction in the objective function value for a given set of hyper-parameters \( \theta \). The EI is mathematically defined as
\[
EI(\theta) = \int_{-\infty}^{L^*} (L^* - L) \frac{l(\theta)}{g(\theta)} \, dL,
\]
calculates the expected improvement by weighting the difference \( L^* - L \) by the ratio 
\[
\frac{l(\theta)}{g(\theta)}
\]
which denotes the relative likelihood of \( \theta \) leading to a better outcome compared to a worse one. The optimal set of hyper-parameters, denoted by \( \theta^* \), is then determined by maximizing the Expected Improvement
\[
\theta^* = \underset{\theta}{\text{argmax}} \, EI(\theta).
\]

By iteratively applying this optimization procedure, the objective is to navigate the hyper-parameter space, focusing the search on regions with a higher probability of improving performance. In the subsequent table and figure, we present the training and validation set accuracies achieved by optimizing CNN classifiers on input data with varying window sizes. We observe that lower window sizes provide superior predictive power, which plateaus with increasing window size. We select 289 trades as the lower threshold for window size, corresponding to approximately ten trading seconds on the AAPL ticker in preliminary calculations. The selection of a lower threshold may still retain a degree of arbitrariness, and exploring different dimensions of window sizes may yield insightful results. \\

\begin{table}[ht!]
\centering
\begin{tabularx}{\linewidth}{X *{2}{X}}
\toprule
Window Size & Training Accuracy (\%) & Validation Accuracy (\%) \\ 
\midrule
289 & 57.74 & 56.57 \\
529 & 54.36 & 51.85 \\
900 & 53.94 & 51.42 \\
1389 & 52.41 & 50.87 \\
1936 & 52.15 & 50.66 \\
\bottomrule
\end{tabularx}
\caption{Training and Validation Accuracy by Window Size.}
\label{table:accuracy_by_window_size}
\end{table}

\begin{figure}[ht!]
    \centering
    \includegraphics[width=8cm, height=6cm]{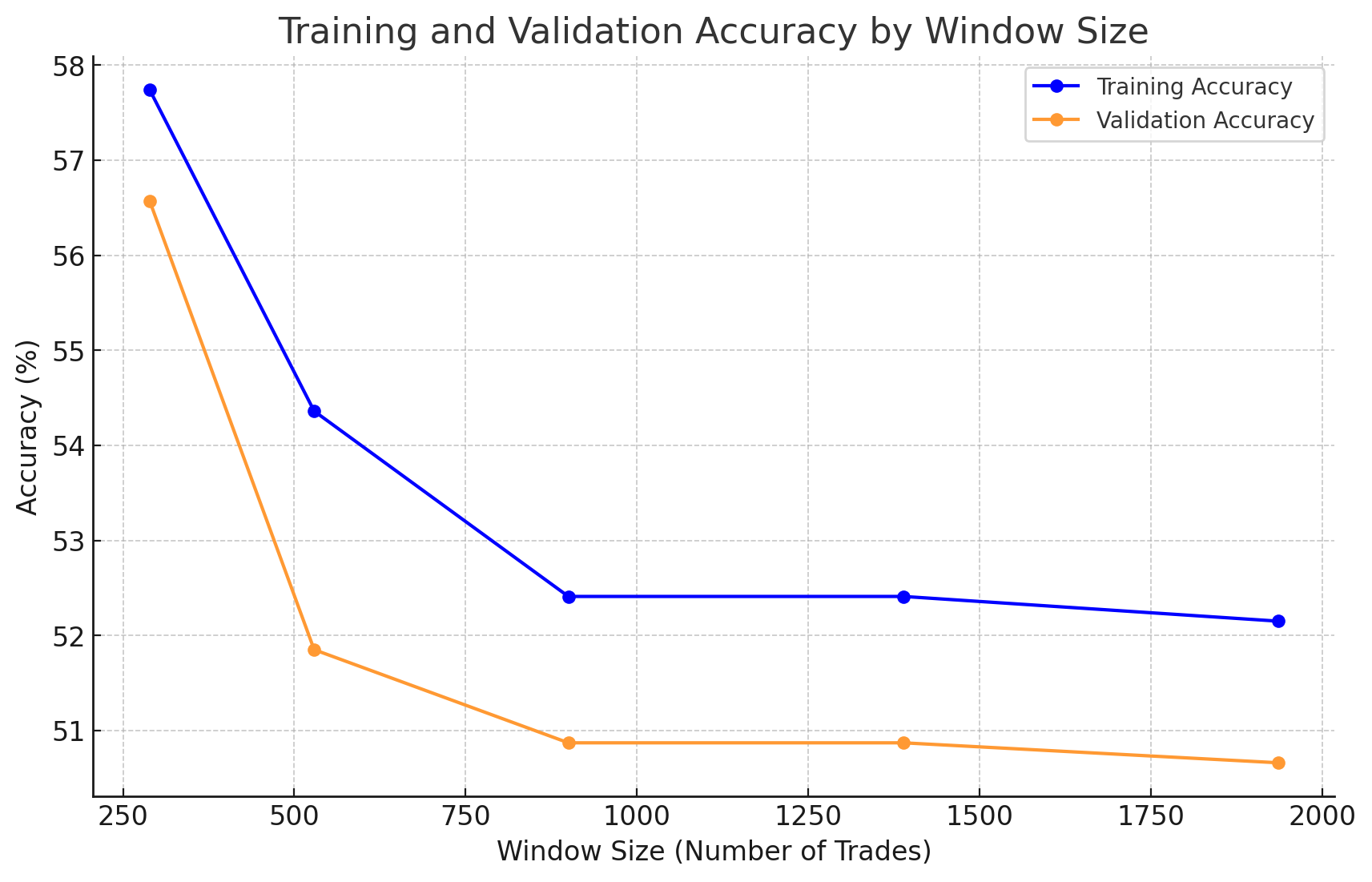}
    \caption{Lower window sizes offer the best predictive power, and this predictability plateaus with increasing window size.}
    \label{fig:fig1}
\end{figure}

\begin{figure}[ht!]
    \centering
    \includegraphics[width=15cm, height=6cm]{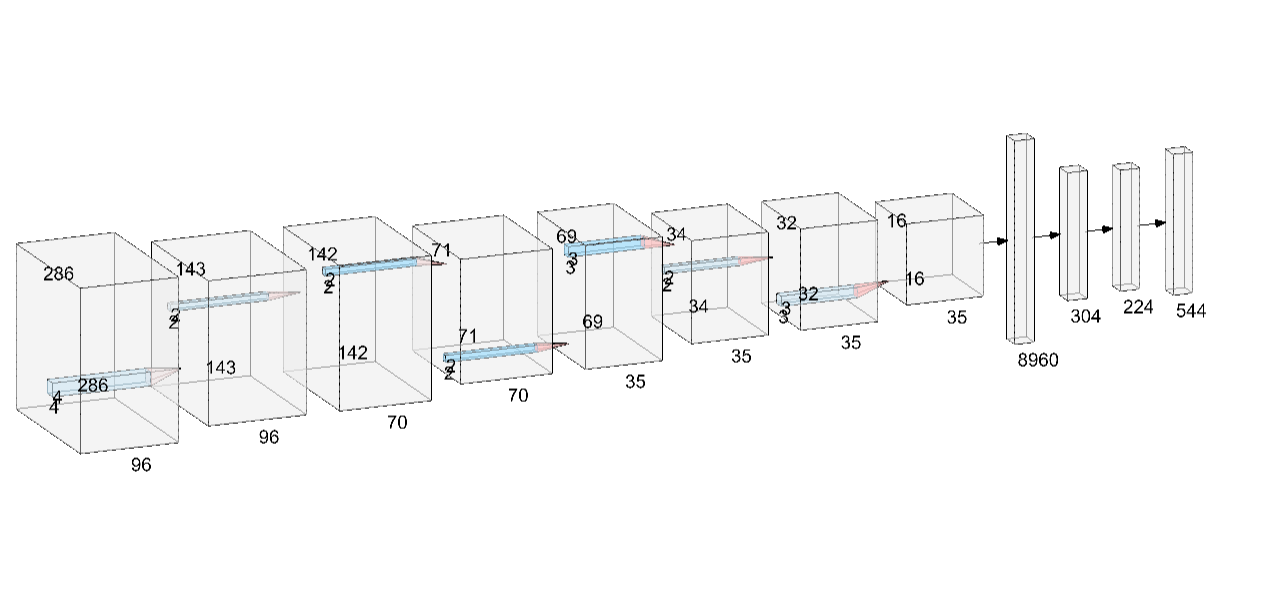}
    \caption{An illustration of the convolutional, pooling and dense layers of our optimal neural network architecture.}
    \label{fig:fig1}
\end{figure}

\begin{table}[ht!]
\centering
\begin{tabularx}{\linewidth}{X *{5}{X}}
\toprule
\textbf{Year} & \textbf{Dataset} & \textbf{Accuracy} & \textbf{Precision} & \textbf{Recall} & \textbf{F1-score} \\ 
\midrule
All & Validation 1 & 0.568 & 0.550 & 0.758 & 0.637 \\
All & Validation 2 & 0.566 & 0.549 & 0.753 & 0.635 \\
2017 & Validation 1  & 0.585 & 0.567 & 0.732 & 0.639 \\ 
2017 & Validation 2  & 0.584 & 0.568 & 0.731 & 0.635 \\ 
2018 & Validation 1  & 0.563 & 0.541 & 0.771 & 0.636 \\ 
2018 & Validation 2 & 0.562 & 0.541 & 0.766 & 0.634 \\ 
2019 & Validation 1   & 0.571 & 0.553 & 0.777 & 0.646 \\ 
2019 & Validation 2 & 0.568 & 0.550 & 0.774 & 0.643 \\ 
2020 & Validation 1   & 0.573 & 0.554 & 0.778 & 0.647 \\ 
2020 & Validation  2 & 0.571 & 0.553 & 0.762 & 0.641 \\ 
2021 & Validation 1   & 0.564 & 0.546 & 0.752 & 0.633 \\ 
2021 & Validation 2 & 0.561 & 0.544 & 0.744 & 0.629 \\ 
\bottomrule
\end{tabularx}
\caption{Model Performance Metrics for two further randomly sampled validation sets (2017-2021).}
\label{table:metrics}
\end{table}

Our optimal architecture comprises four convolutional layers and three dense layers. The first convolutional layer consists of 96 filters with a kernel size of 4, the second has 70 filters with a kernel size of 2, and the third contains 35 filters with a kernel size of 3, all using 'He normal' initializer (used to better manage gradients,  \cite{he2015delving}). The dense layers have 304, 224, and 544 units, respectively, with dropout (\cite{JMLR:v15:srivastava14a}) rates of 0.2729, 0.3348, and 0.1340. The learning rate is set at 0.0010427, and gradient clipping (\cite{zhang2020gradient}) \footnote{Gradient clipping is a technique employed in training neural networks to prevent the occurrence of excessively large gradients, which can lead to numerical instability and hinder the convergence of the optimization algorithm. During the back-propagation process, gradients are computed to update the parameters of the model. However, in certain scenarios, especially in deep networks these gradients can become excessively large, a phenomenon known as the "exploding gradient" problem. This can lead to issues such as overflow, null values, and erratic updates that can cause the training process to diverge. To mitigate this issue, gradient clipping imposes a threshold on the gradients, ensuring that their magnitude does not exceed a specified maximum value.} is applied with a value of 1.0306. The training accuracy of this configuration is 57.74\%, and the validation accuracy is 56.57\% at a size of 289 trades per window. We tested our predictor on one further held-out sample, and report the results from that test in Table 3. Our predictor has similar accuracies in both held-out sets, and these accuracies are stable across the time period of our sample. 

\subsection{Inference Problem: Identifying Influential Trades}

In the previous section, we were able to configure CNN predictors to achieve robust prediction accuracies. The neural network predictor \( f(\cdot; \theta) \) thus contains information relating to how specific inputs are systematically associated with the direction of future price movements. In this section, we describe our approach towards extracting attributes of input trades that most significantly influence the Neural Network's predictions. To this end, we follow an approach based on the seminal work of \cite{simonyan2014deep}. Several factors were important to consider in deciding which candidate approaches were most appropriate to our study. First, in contrast to approaches that examine the importance of input features primarily in the context of the overall model (for example in Random Forests, see \cite{Breiman2001RandomForests}), our application requires us to compute influence scores of input attributes for each observation. \\

The widely popular Shapely Values related approaches of \cite{Lundberg2017}, \cite{sundararajan2017axiomatic} are grounded in theory and can be used to compute influence scores of attributes for each input observation, but are computationally intensive and require definitions for baseline values. Other popular approaches that rely on building surrogate models (for example the seminal approach of \cite{ribeiro2016why}) may suffer in performance in applications where decision boundaries may be especially irregular, as is likely in our case. Still newer approaches that improve on  \cite{simonyan2014deep}, (for example \cite{zhou2015learning}; \cite{Selvaraju_2019};\cite{sundararajan2017axiomatic}; \cite{smilkov2017smoothgrad}; \cite{srinivas2019fullgradient}) may be candidates for future exploration in this area. \\

Having previously trained our predictor, we obtain a function \( f(.; \theta) \) that accurately estimates the direction of future price movements \( y_{t+1} \) given an input tensor of trade attributes \( \mathbf{X}^{(t)} \). The predicted probability of \( y_{t+1} = 1 \) from the neural network is \( \hat{y}_{t+1} \). We present the fully trained Neural Network with its own prediction score \( \hat{y}_{t+1} \) and query the influence of each input feature \( x_{ijc}^{(t)} \) on \( \hat{y}_{t+1} \), where \( i \) and \( j \) index the trade position within the window, and \( c \) indexes the channel. Suppose \( f(\cdot; \theta) \) were linear, expressed as:
\[ \hat{y}_{t+1} = f(\mathbf{X}^{(t)}; \theta) = \sum_{i,j,c} \theta_{ijc} \cdot x_{ijc}^{(t)} + b. \]
Then, the influence of each attribute on \( \hat{y}_{t+1} \) would be straightforward to determine and given by:
\[ \mathbf{w}_{ijc} = \frac{\partial f(\mathbf{X}^{(t)}; \theta)}{\partial x_{ijc}^{(t)}} = \theta_{ijc} \]
where \( \mathbf{w}_{ijc} \) represents the influence score of the attribute in position \( (i, j, c) \) of the input tensor \( \mathbf{X}^{(t)} \). However, \( f(\cdot; \theta) \) is non-linear, and thus the influence of the input attributes on the final predicted score must be approximated numerically. We compute a first-order Taylor approximation for each input tensor \( \mathbf{X}^{(t)} \) of dimension \( m \times m \times 35 \) and a predicted probability \( \hat{y}_{t+1} \), yielding a tensor of influence scores \( \mathbf{w}_{ijc} \). To aggregate the final influence score for each trade, we sum the absolute values of the influence scores of all attributes (across all channels) for that trade:
\[ \text{Final Influence Score} = \sum_{c=1}^{35} |\mathbf{w}_{ijc}|. \]
Larger aggregate scores for a trade in a given window indicates that the trade had a greater impact on the optimized predictor's prediction of future price movements, while smaller scores suggest that the trade in question exerted a negligible influence on the optimized predictor's prediction of future price movements.

\subsection{Extracting Attributes of Influential Trades}

In this section, we focus on utilizing the influence scores calculated from the previous section for each trade over the full sample to identify characteristics of influential trades. For every trade \(i\) in trading window \(t\), we calculate its influence score \(w_{it}\). In addition to the influence score \(w_{it}\) for each trade \(i\) in window \(t\), we create our regression variables, shown in Table 7, based on the trade's characteristics. We estimate the relative effect on influence scores using the following regression equation:

\begin{align}
    w_{it} = \mathbf{T}_{it} + \mathbf{Y}_{it} + \mathbf{E}_{it} + \mathbf{T*Y}_{it} + \mathbf{T*E}_{it} + \mathbf{Y*E}_{it} + \mathbf{Td*T}_{it} + \mathbf{Td*Y}_{it} + \mathbf{Td*E}_{it} + \epsilon.
\end{align}

In this regression equation,
 \(\mathbf{T}_{it}\) represents the trade size Indicators.
 \(\mathbf{Y}_{it}\) represents the year dummies.
 \(\mathbf{E}_{it}\) represents the exchange dummies.
 \(\mathbf{T*Y}_{it}\), \(\mathbf{T*E}_{it}\), \(\mathbf{Y*E}_{it}\) represent interaction terms for Trade Size-Year, Trade Size-Exchange, and Year-Exchange, respectively.
 \(\mathbf{Td*T}_{it}\), \(\mathbf{Td*Y}_{it}\), \(\mathbf{Td*E}_{it}\) represent interaction terms for Time Difference-Trade Size, Time Difference-Year, and Time Difference-Exchange, respectively.
 \(\epsilon\) is the error term. Standard errors are clustered at the window level. Reference variables for each category of dummies in the model are detailed in the table below.

\begin{table}[ht!]
\centering
{\small
\begin{tabularx}{\linewidth}{>{\centering\arraybackslash}X >{\centering\arraybackslash}X}
\toprule
Reference Variable & Description \\
\midrule
Round\_lot & Reference for Trade Size Indicators \\
2017 & Reference for Year Dummies \\
NYSE & Reference for Exchange Dummies \\
\bottomrule
\end{tabularx}
}
\caption{Description of Reference Variables in Regression Equation}
\label{table:3}
\end{table}

The regression equation is estimated separately for the following three conditions:
     (1) For the full training (in sample) set,
    (2) For the training (in sample) set containing only positive price predictions (predicted output = 1),
     (3) For the training (in sample) set containing only negative price predictions (predicted output = 0).

\section{Results and Discussion} \label{sec:result}

In Table 7 and Figure 5, we report and visualize summary statistics and the distributions of influence scores of trades with different attributes. The average trade has an influence score of 52.90, with a standard deviation of 70.70. The maximum influence score observed is 1601.27, and minimum score is 0. The median influence score is 27.69. Figure 4 shows that the distribution of influence scores has a long right tail, suggesting that a minority of highly predictive "bell-weather" trades are dispersed in a large majority of "vanilla" trades in our sample. The average influence score for trades in windows where future negative action is predicted is 52.81, with a sample standard deviation of 65.69. For trades in windows where future positive price movements are predicted, the average score is slightly higher at 52.95, with a sample standard deviation of 72.98. The average influence score for stock trades is 54.59, while the average score for ETF trades is notably lower at 39.89. The sample standard deviation for stocks stands at 71.44, compared to a lower sample standard deviation of 63.29 for ETFs. The maximum influence score observed is marginally higher for stocks at 1601.27, compared to 1550.34 for ETFs. \\

The influence scores vary based on different trade size categories. Round lot trades display the highest average influence score, at 53.23. Odd-lot trades and trades between sizes 101-1000 show similar influence-score averages of 46.12 and 46.13 respectively. Trades between sizes 1001-10000 and greater than 10000 shares have considerably lower and decreasing average influence scores of 34.57 and 29.98 respectively. The sample influence-score standard deviations for trades in these trade-size categories in increasing order of magnitude are 64.99, 70.11, 64.99, 52.79, and 45.75, respectively. The average influence scores for trades conducted in each year of our sample are as follows: 48.97 for 2017, 51.64 for 2018, 54.08 for 2019, 51.86 for 2020, and 54.77 for 2021. The sample influence-score standard deviations for these years are 65.87, 69.90, 71.21, 69.43, and 72.82 respectively, indicating a slight increase in variability over time. The yearly breakdown reveals an overall upward trend in average influence scores, with a slight dip in 2020. Our preliminary analysis also reveals varying average influence scores for trades conducted at different venues.\\

We then conducted an analysis of the correlation between different attributes of trades in our sample. The estimates are shown in Table 8. The positive correlation (0.12) between ETF trades and the 101 to 1000 lot size category suggests that trades of this size range are associated more with ETFs than with stocks in our sample. The negative correlation (-0.11) between ETF trades and odd lot sizes indicates that odd-lot trades are slightly more prevalent in stocks than ETFs in our sample. The positive correlation (0.03, 0.24) between odd lot trades and the years 2020, 2021 in our sample might likely reflect the increased participation of retail traders overall. Certain exchanges, such as MIAX, show a positive correlation with larger trade sizes (greater than 10000), potentially indicating their role in facilitating institutional trading activities, due to the specific market structure and order matching mechanisms of these exchanges. On the other hand, exchanges like NASDAQ OMX BX show a propensity for smaller, non-ETF trades, evidenced by the positive correlation with round lot trades and negative correlation with ETF trades. The full set of correlation estimates is reported in Table 8.  \\

Next, we use the influence scores extracted from our prediction-inference framework to discuss the characteristics of trades that most predict future market movements. We use a randomly extracted fraction of our trades data, 28.9 million trades to estimate our regression coefficients and standard errors because of memory constraints. We show in our Tables estimates from model 1, and progressively add regressors. To begin, we include basic trade characteristics and the year the trade was conducted. Then, we progressively add trade size, size-year interactions, volume-year interactions and venue fixed-effects. The estimates are shown from Tables 7-11. Plots of regression estimates from our full model, from Table 11, are shown in Figures 5-10.  In all regression tables, column 1 shows the estimates from the full sample, and columns 2 and 3 show estimates extracted from trading windows that were extracted from windows that were assigned positive and negative predictions respectively. All standard errors are clustered at the window-level. All estimates were extracted from trading-windows originally in our validation set. We separately conducted a robustness exercise to ensure that these estimates did not vary much across 1) different samples of trades 2) using the training set of windows rather than the validation windows to construct regression estimates. It is to be noted that while these estimates are from trades in our sample and during the period 2017-2021, our model permits estimates from other timelines or for a different set of trading windows to be different from these results.  \\

Our analysis revealed that trades of size less than 100 shares (odd lot trades) exhibited a slightly higher predictability compared to trades of 100 shares (round lot trades), with an effect size of 0.58 (p < 0.10). Interestingly, this effect was more pronounced for windows predicting positive market movements, where odd lot trades demonstrated greater predictability in forecasting future price movements, with an effect size of 1.06 (p < 0.001). However, the effect was not significant in predicted negative windows, with an effect size of 0.37. Trades between sizes of 101-1000 shares exhibited less predictability than round lot trades, with an effect size of -5.14 (p < 0.001). This effect was statistically significant and consistent across both windows predicting future positive movements, with an effect size of -5.14 (p < 0.001), and those predicting future negative price movements, with an effect size of -4.81 (p < 0.001). Trades of sizes between 1001 and 9999 shares were found to be less influential than round lot trades in predicting future market movements, with an effect size of -14.95 (p < 0.001). In windows predicting future negative movements, the effect was significant, with effect sizes of -15.83 (p < 0.001) for predicted positive windows and -11.43 (p < 0.001) for predicted negative windows. Finally, our analysis indicated that trades of sizes greater than or equal to 10,000 shares were substantially less influential compared to round lot trades in predicting future market movements. This negative effect was significant, large, and persisted across both windows predicting future positive movements, with an effect size of -20.28 (p < 0.001), and those predicting future negative price movements, with an effect size of -21.47 (p < 0.001) for predicted positive windows and -15.15 (p < 0.001) for predicted negative windows. \\

For odd lot trades, we observe varying effects across different years. In 2018, there is no significant impact of odd lot trades on overall predictability, but a positive effect (1.09, p < 0.05) is observed in predicted positive windows, suggesting that odd lot trades in 2018 offered more information during positive future market movements. However, a strong negative effect (-3.17, p < 0.001) is noted in predicted negative windows, indicating that odd lot trades were particularly predictive of negative market movements in 2018. In 2019, odd lot trades consistently show a negative impact across all models, with a significant effect in overall trades (-1.26, p < 0.001) and predicted positive windows (-1.22, p < 0.001). The year 2020 presents a change, with a substantial positive effect in both overall trades (5.51, p < 0.001) and predicted positive windows (7.03, p < 0.001), showing an increased predictive power of odd lot trades during this period. In 2021, the positive impact persists in predicted positive windows (4.15, p < 0.001), although with a slightly reduced magnitude compared to 2020. \\

For trades of size between 101 and 1000 shares, we observe consistent positive effects in 2018 across all models, with a notable impact in overall trades (1.87, p < 0.001) and predicted positive windows (1.34, p < 0.001). The year 2019 shows a moderate positive effect, particularly in overall trades (0.81, p < 0.001) and predicted negative windows (0.79, p < 0.05). However, in 2020, a negative effect is observed in predicted positive windows (-3.10, p < 0.001), suggesting reduced predictability, while a strong positive effect is noted in predicted negative windows (5.10, p < 0.001). In 2021, a significant negative impact is seen across all models, with the most pronounced effect in predicted positive windows (-2.48, p < 0.001). Trades of size between 1001 and 9999 shares exhibit positive effects in 2018 and 2019, with the highest impact observed in 2018 for overall trades (4.87, p < 0.001) and predicted positive windows (3.99, p < 0.001). The year 2020 shows a mixed effect, with a positive impact in overall trades (4.00, p < 0.001) but a negative effect in predicted positive windows (-1.45). In 2021, a significant negative effect is observed across all models, with the largest impact in predicted positive windows (-6.13, p < 0.001). For trades greater than or equal to 10,000 shares, the interactions reveal a strong positive effect in 2018 and 2019, particularly in overall trades (8.08, p < 0.001 for 2018) and predicted positive windows (5.82, p < 0.001 for 2019). The year 2020 exhibits the highest positive impact in predicted negative windows (15.52, p < 0.001), indicating increased predictability during this volatile period. In contrast, 2021 shows a significant negative effect, especially in predicted positive windows (-8.17, p < 0.001). \\

ETF trades show a significant negative effect across all models, indicating that trades involving ETFs are generally less predictive of future market movements compared to trades involving stocks in our sample. Specifically, the coefficient for overall trades is -11.76 (p < 0.001), a substantial reduction given the full sample average influence score of 52.99. This effect is less pronounced but still significant in predicted positive windows, with an effect size of -6.47 (p < 0.001), indicating that ETF trades are less predictive of positive future market movements. The negative impact is most pronounced in predicted negative windows, with an effect size of -29.14 (p < 0.001), suggesting that ETF trades are particularly less informative than stock trades in predicting negative future market movements. The effect sizes for ETFs is marginally reduced further when trade sizes are removed from the model, suggesting that varying trade size distributions between stock and ETF trades may explain a small portion of the reduced predictability offered by ETF trades as compared to stocks about future price movements. A study of other factors that could potentially explain the reduced predictability offered by ETFs as compared to stocks may be valuable extensions to our present work. \\

In the overall model, the coefficient for the variable indicating the natural logarithm of the time difference in nanoseconds between the focal trade and the trade trade immediately preceding it is 0.03, and is not statistically significant (p > 0.05). In the model for predicted positive movements, the coefficient is 0.10 (p < 0.001), indicating that a longer time since the last trade may be associated with a higher influence score, suggesting that trades that occur after a period of inactivity may carry more predictive information for positive market movements in our sample. Conversely, in the model for predicted negative movements, the coefficient is -0.09 (p < 0.001), suggesting that a longer time since the last trade is associated with a lower influence score, indicating that such trades may be less informative for predicting negative market movements in our sample. The interaction terms and the year dummies provide further insights. The interaction with 2018 is positive and significant in the overall model (0.07, p < 0.001) and in the model for predicted negative movements (0.10, p < 0.01), suggesting that the impact of time since the last trade on influence scores was more pronounced in 2018, particularly for predicting negative market movements. The interaction with 2020 is also positive and significant across all models, with the largest effect observed in the model for predicted negative movements (0.34, p < 0.001), indicating that the time elapsed since the last trade was particularly informative for predicting negative market movements in the year of 2020. On the other hand, the interaction with 2021 is negative and significant in the model for predicted positive movements (-0.22, p < 0.001) in our sample, suggesting that in 2021, a longer time since the last trade was associated with lower predictability for positive market movements. We included the variable indicating the natural logarithm of the time difference in nanoseconds between the focal trade and the trade trade immediately preceding it to be a measure of the high/low volume trading environment the trade occurs in. Other choices of measures that could better capture this aspect of the trading environment may provide further utility when used in a similar model. \\

We then compare the influence scores across various exchanges with the New York Stock Exchange (NYSE) as the reference category. Controlling for all our other attributes, trades conducted at the NASDAQ OMX BX and Cboe BYX show the most pronounced negative effects among the venues analyzed, in our sample. For NASDAQ OMX BX, the coefficients for the overall trades, predicted positive movements, and predicted negative movements are -15.76 (p < 0.001), -13.60 (p < 0.001), and -21.76 (p < 0.001), respectively. Similarly, for Cboe BYX, the corresponding coefficients are -15.36 (p < 0.001), -13.77 (p < 0.001), and -20.44 (p < 0.001). These significant negative coefficients indicate that trades conducted at these venues may carry less predictive information for future market movements when compared with trades conducted at the NYSE. On the other end, controlling for all our other attributes, trades conducted at MIAX stand out with a positive effect. The coefficients for trades on MIAX are 11.31 (p < 0.001) for overall trades, 10.31 (p < 0.001) for predicted positive movements, and 12.44 (p < 0.001) for predicted negative movements. This positive association suggests that trades on MIAX have higher influence scores compared to the NYSE in our sample. The full set of estimates relating to influence scores of trades conducted at different venues are shown in Table 11, and in Figures 8, 9 and 10. \\

\section{Discussion and Future Directions} \label{sec:conclusion}

In this paper, we demonstrated an approach to identify individual trades that influence a successful Black-Box Neural Network's prediction of future market movements. We further submitted an optimization-prediction-interpretation framework, where the best predictor among a class of predictors was identified and interpreted. Our approach to identifying trades that most influence an optimized predictor's prediction of future market movements does not require or assume a model of the trading process. This means that hypotheses generated from theoretical treatments to the attributes of trades that most influence future market movements could be tested using our approach. Our approach allows us to monitor the attributes of trades that most influence future market movements over time and by venue, among other attributes. This information could inform policy/strategy for policy-makers, traders, and exchanges. Extending our analysis, it is possible to study the attributes of trades that most influence future market movements with varying window sizes. This analysis could provide useful information on which trade attributes best inform short vs long-term predictions. Similarly, a study of differences in emphasis on specific attributes of trades between successful and unsuccessful models may be useful in building better prediction models of future market movements. Our analysis could be extended to markets for options, bonds, and other instruments.  \\

 In this work, we used a window of individual trades to predict future price movements. The same analysis could be performed on time or volume bars to find the attributes of points in time that are most influential in determining future price movements. If, instead of time/volume bars, established microstructure variables are provided as inputs, the approach allows the identification of which microstructure variables might be more relevant for predicting future price movements on a case-by-case basis. We identified trade timing, trade size, venue, measures of trading volume, year, and the interactions of these variables as attributes of trades that explain the influence of these trades. It is an interesting future direction of research to explore a more extensive set of factors that could explain the influence of trades, including the size and location of spread and measures created from multiple levels of the order book. This potential for future research directions is exciting and could significantly contribute to the field of financial market analysis. The data used in this work relates to a period of five years and stocks/ETFs. It might be interesting to include data from a longer period of time, and possibly further analyse temporal changes in factors that affect the influence of individual trades on future market movements. It might be interesting to ask similar questions across assets than the present within-asset analysis. \\

 Some predictors may be better suited to exploit certain aspects of input data (for example, CNNs have been observed to have an advantage in detecting patterns). Empirically, it may be helpful to understand differences in attributes of trades emphasized by different classes of predictors. Doing so might provide useful information to build better prediction frameworks. Finally, our framework could provide a template for introducing further robustness in analyses that seek to use prediction models for downstream analyses by helping identify the optimal predictor from a class of models.   

\newpage
\bibliographystyle{plainnat}
\bibliography{references.bib}

\begin{thebibliography}{52}
\providecommand{\natexlab}[1]{#1}
\providecommand{\url}[1]{\texttt{#1}}
\expandafter\ifx\csname urlstyle\endcsname\relax
  \providecommand{\doi}[1]{doi: #1}\else
  \providecommand{\doi}{doi: \begingroup \urlstyle{rm}\Url}\fi

\bibitem[Akiba et~al.(2019)Akiba, Sano, Yanase, Ohta, and
  Koyama]{akiba2019optuna}
Takuya Akiba, Shotaro Sano, Toshihiko Yanase, Takeru Ohta, and Masanori Koyama.
\newblock Optuna: A next-generation hyperparameter optimization framework.
\newblock In \emph{Proceedings of the 25th ACM SIGKDD international conference
  on knowledge discovery \& data mining}, pages 2623--2631, 2019.

\bibitem[Athey and Imbens(2019)]{athey2019machine}
Susan Athey and Guido~W Imbens.
\newblock Machine learning methods economists should know about.
\newblock \emph{Annual Review of Economics}, 11\penalty0 (1):\penalty0
  685--725, 2019.

\bibitem[Barclay and Warner(1993)]{RePEc:eee:jfinec:v:34:y:1993:i:3:p:281-305}
Michael~J. Barclay and Jerold~B. Warner.
\newblock Stealth trading and volatility: Which trades move prices?
\newblock \emph{Journal of Financial Economics}, 34\penalty0 (3):\penalty0
  281--305, 1993.
\newblock URL
  \url{https://EconPapers.repec.org/RePEc:eee:jfinec:v:34:y:1993:i:3:p:281-305}.

\bibitem[Bergstra et~al.(2011)Bergstra, Bardenet, Bengio, and
  K\'{e}gl]{NIPS2011_86e8f7ab}
James Bergstra, R\'{e}mi Bardenet, Yoshua Bengio, and Bal\'{a}zs K\'{e}gl.
\newblock Algorithms for hyper-parameter optimization.
\newblock In J.~Shawe-Taylor, R.~Zemel, P.~Bartlett, F.~Pereira, and K.Q.
  Weinberger, editors, \emph{Advances in Neural Information Processing
  Systems}, volume~24. Curran Associates, Inc., 2011.
\newblock URL
  \url{https://proceedings.neurips.cc/paper_files/paper/2011/file/86e8f7ab32cfd12577bc2619bc635690-Paper.pdf}.

\bibitem[Bianchi et~al.(2020)Bianchi, Buyl, and Vanduffel]{bianchi2020}
Daniele Bianchi, S.~Buyl, and S.~Vanduffel.
\newblock Deep learning in portfolio allocation.
\newblock \emph{Journal of Financial Economics}, 138\penalty0 (3):\penalty0
  442--456, 2020.

\bibitem[Bishop(2006)]{Bishop2006Pattern}
Christopher~M. Bishop.
\newblock \emph{Pattern Recognition and Machine Learning}.
\newblock Springer, 2006.

\bibitem[Breiman(2001)]{Breiman2001RandomForests}
Leo Breiman.
\newblock Random forests.
\newblock \emph{Machine Learning}, 45:\penalty0 5--32, 2001.
\newblock \doi{10.1023/A:1010933404324}.
\newblock URL \url{https://doi.org/10.1023/A:1010933404324}.

\bibitem[Campigli et~al.(2022)Campigli, Bormetti, and
  Lillo]{campigli2023measuring}
Francesco Campigli, Giacomo Bormetti, and Fabrizio Lillo.
\newblock Measuring price impact and information content of trades in a
  time-varying setting.
\newblock \emph{arXiv preprint arXiv:2212.12687}, 2022.

\bibitem[Chakravarty(2001)]{CHAKRAVARTY2001289}
Sugato Chakravarty.
\newblock Stealth-trading: Which traders’ trades move stock prices?
\newblock \emph{Journal of Financial Economics}, 61\penalty0 (2):\penalty0
  289--307, 2001.
\newblock ISSN 0304-405X.
\newblock \doi{https://doi.org/10.1016/S0304-405X(01)00063-0}.
\newblock URL
  \url{https://www.sciencedirect.com/science/article/pii/S0304405X01000630}.

\bibitem[Chen et~al.(2020)Chen, Pelger, and Zhu]{chen2020}
Li~Chen, Markus Pelger, and Jason Zhu.
\newblock Deep learning in asset pricing.
\newblock \emph{Journal of Financial Economics}, 140\penalty0 (1):\penalty0
  1--30, 2020.

\bibitem[Chinco et~al.(2019)Chinco, Clark-Joseph, and Ye]{chinco2018}
Alex Chinco, Adam Clark-Joseph, and Mao Ye.
\newblock Sparse signals in the cross-section of returns.
\newblock \emph{The Journal of Finance}, 74\penalty0 (1):\penalty0 449--492,
  2019.

\bibitem[de~Prado(2018)]{deprado2018}
Marcos~Lopez de~Prado.
\newblock \emph{Advances in Financial Machine Learning}.
\newblock John Wiley \& Sons, 2018.

\bibitem[Dufour and Nguyen(2012)]{RePEc:taf:eurjfi:v:18:y:2012:i:9:p:841-864}
Alfonso Dufour and Minh Nguyen.
\newblock Permanent trading impacts and bond yields.
\newblock \emph{The European Journal of Finance}, 18\penalty0 (9):\penalty0
  841--864, 2012.
\newblock URL
  \url{https://EconPapers.repec.org/RePEc:taf:eurjfi:v:18:y:2012:i:9:p:841-864}.

\bibitem[Easley et~al.(1997)Easley, Kiefer, and
  O'Hara]{2f4daf45-d507-3e93-8916-77093703379f}
David Easley, Nicholas~M. Kiefer, and Maureen O'Hara.
\newblock One day in the life of a very common stock.
\newblock \emph{The Review of Financial Studies}, 10\penalty0 (3):\penalty0
  805--835, 1997.
\newblock ISSN 08939454, 14657368.
\newblock URL \url{http://www.jstor.org/stable/2962204}.

\bibitem[Easley et~al.(2019)Easley, de~Prado, O'Hara, and
  Zhang]{easley2019microstructure}
David Easley, Marcos~Lopez de~Prado, Maureen O'Hara, and Zhibai Zhang.
\newblock Microstructure in the machine learning age.
\newblock \emph{The Review of Financial Studies}, 34\penalty0 (7):\penalty0
  3316--3363, 2019.

\bibitem[Easley et~al.(2021)Easley, de~Prado, O’Hara, Zhang, and
  Jiang]{RePEc:oup:rfinst:v:34:y:2021:i:7:p:3316-3363.}
David Easley, Marcos~López de~Prado, Maureen O’Hara, Zhibai Zhang, and Wei
  Jiang.
\newblock {Microstructure in the Machine Age [The risk of machine learning]}.
\newblock \emph{The Review of Financial Studies}, 34\penalty0 (7):\penalty0
  3316--3363, 2021.
\newblock URL
  \url{https://ideas.repec.org/a/oup/rfinst/v34y2021i7p3316-3363..html}.

\bibitem[Engle and Patton(2004)]{RePEc:eee:finmar:v:7:y:2004:i:1:p:1-25}
Robert Engle and Andrew Patton.
\newblock Impacts of trades in an error-correction model of quote prices.
\newblock \emph{Journal of Financial Markets}, 7\penalty0 (1):\penalty0 1--25,
  2004.
\newblock URL
  \url{https://EconPapers.repec.org/RePEc:eee:finmar:v:7:y:2004:i:1:p:1-25}.

\bibitem[Feng et~al.(2019)Feng, He, and Polson]{feng2019}
Guanhao Feng, He~He, and Nicholas~G. Polson.
\newblock Deep learning for asset pricing.
\newblock \emph{Journal of Financial and Quantitative Analysis}, 54\penalty0
  (3):\penalty0 1187--1224, 2019.

\bibitem[Glosten and Milgrom(1985)]{GLOSTEN198571}
Lawrence~R. Glosten and Paul~R. Milgrom.
\newblock Bid, ask and transaction prices in a specialist market with
  heterogeneously informed traders.
\newblock \emph{Journal of Financial Economics}, 14\penalty0 (1):\penalty0
  71--100, 1985.
\newblock ISSN 0304-405X.
\newblock \doi{https://doi.org/10.1016/0304-405X(85)90044-3}.
\newblock URL
  \url{https://www.sciencedirect.com/science/article/pii/0304405X85900443}.

\bibitem[Goodfellow et~al.(2016)Goodfellow, Bengio, and
  Courville]{Goodfellow-et-al-2016}
Ian Goodfellow, Yoshua Bengio, and Aaron Courville.
\newblock \emph{Deep Learning}.
\newblock MIT Press, 2016.
\newblock \url{http://www.deeplearningbook.org}.

\bibitem[Gu et~al.(2020)Gu, Kelly, and Xiu]{gu2020}
Shihao Gu, Bryan Kelly, and Dacheng Xiu.
\newblock Empirical asset pricing via machine learning.
\newblock \emph{Journal of Financial Economics}, 142\penalty0 (2):\penalty0
  701--725, 2020.

\bibitem[Hansch and Choe(2005)]{article}
Oliver Hansch and Hyuk Choe.
\newblock Which trades move stock prices in the internet age?
\newblock \emph{SSRN Electronic Journal}, 03 2005.
\newblock \doi{10.2139/ssrn.676986}.

\bibitem[Hasbrouck(1991)]{https://doi.org/10.1111/j.1540-6261.1991.tb03749.x}
Joel Hasbrouck.
\newblock Measuring the information content of stock trades.
\newblock \emph{The Journal of Finance}, 46\penalty0 (1):\penalty0 179--207,
  1991.
\newblock \doi{https://doi.org/10.1111/j.1540-6261.1991.tb03749.x}.
\newblock URL
  \url{https://onlinelibrary.wiley.com/doi/abs/10.1111/j.1540-6261.1991.tb03749.x}.

\bibitem[Hasbrouck(1995)]{https://doi.org/10.1111/j.1540-6261.1995.tb04054.x}
Joel Hasbrouck.
\newblock One security, many markets: Determining the contributions to price
  discovery.
\newblock \emph{The Journal of Finance}, 50\penalty0 (4):\penalty0 1175--1199,
  1995.
\newblock \doi{https://doi.org/10.1111/j.1540-6261.1995.tb04054.x}.
\newblock URL
  \url{https://onlinelibrary.wiley.com/doi/abs/10.1111/j.1540-6261.1995.tb04054.x}.

\bibitem[Hastie et~al.(2009)Hastie, Tibshirani, and
  Friedman]{Hastie2009Elements}
Trevor Hastie, Robert Tibshirani, and Jerome Friedman.
\newblock \emph{The Elements of Statistical Learning: Data Mining, Inference,
  and Prediction}.
\newblock Springer Science \& Business Media, 2009.

\bibitem[Hautsch and Huang(2012)]{RePEc:eee:dyncon:v:36:y:2012:i:4:p:501-522}
Nikolaus Hautsch and Ruihong Huang.
\newblock The market impact of a limit order.
\newblock \emph{Journal of Economic Dynamics and Control}, 36\penalty0
  (4):\penalty0 501--522, 2012.
\newblock URL
  \url{https://EconPapers.repec.org/RePEc:eee:dyncon:v:36:y:2012:i:4:p:501-522}.

\bibitem[He et~al.(2015)He, Zhang, Ren, and Sun]{he2015delving}
Kaiming He, Xiangyu Zhang, Shaoqing Ren, and Jian Sun.
\newblock Delving deep into rectifiers: Surpassing human-level performance on
  imagenet classification.
\newblock In \emph{Proceedings of the IEEE International Conference on Computer
  Vision}, pages 1026--1034, 2015.

\bibitem[Heaton et~al.(2017)Heaton, Polson, and Witte]{heaton2017}
J.B. Heaton, N.G. Polson, and J.H. Witte.
\newblock Deep learning for finance: Deep portfolios.
\newblock \emph{Applied Stochastic Models in Business and Industry},
  33\penalty0 (1):\penalty0 3--12, 2017.

\bibitem[Kelly and Xiu(2023)]{kelly2023}
Bryan Kelly and Dacheng Xiu.
\newblock Financial networks and asset returns.
\newblock \emph{Review of Financial Studies}, 2023.
\newblock \doi{10.1093/rfs/hhab014}.

\bibitem[Krauss et~al.(2017)Krauss, Do, and Huck]{krauss2017}
Christopher Krauss, X.~Do, and Nils Huck.
\newblock Deep neural networks, gradient-boosted trees, random forests:
  Statistical arbitrage on the s\&p 500.
\newblock \emph{European Journal of Operational Research}, 259\penalty0
  (2):\penalty0 689--702, 2017.

\bibitem[Kyle(1985)]{e2104c33-2a9d-3e4e-be9a-35cae4feb70d}
Albert~S. Kyle.
\newblock Continuous auctions and insider trading.
\newblock \emph{Econometrica}, 53\penalty0 (6):\penalty0 1315--1335, 1985.
\newblock ISSN 00129682, 14680262.
\newblock URL \url{http://www.jstor.org/stable/1913210}.

\bibitem[LeCun et~al.(1989)LeCun, Boser, Denker, Henderson, Howard, Hubbard,
  and Jackel]{6795724}
Y.~LeCun, B.~Boser, J.~S. Denker, D.~Henderson, R.~E. Howard, W.~Hubbard, and
  L.~D. Jackel.
\newblock Backpropagation applied to handwritten zip code recognition.
\newblock \emph{Neural Computation}, 1\penalty0 (4):\penalty0 541--551, 1989.
\newblock \doi{10.1162/neco.1989.1.4.541}.

\bibitem[Li et~al.(2020)Li, Yang, Peng, and
  Liu]{DBLP:journals/corr/abs-2004-02806}
Zewen Li, Wenjie Yang, Shouheng Peng, and Fan Liu.
\newblock A survey of convolutional neural networks: Analysis, applications,
  and prospects.
\newblock \emph{CoRR}, abs/2004.02806, 2020.
\newblock URL \url{https://arxiv.org/abs/2004.02806}.

\bibitem[Lundberg and Lee(2017)]{Lundberg2017}
Scott~M. Lundberg and Su-In Lee.
\newblock A unified approach to interpreting model predictions.
\newblock In \emph{Advances in Neural Information Processing Systems}, 2017.
\newblock URL
  \url{https://papers.nips.cc/paper/2017/hash/8a20a8621978632d76c43dfd28b67767-Abstract.html}.

\bibitem[O'Hara et~al.(2014)O'Hara, Yao, and
  Ye]{https://doi.org/10.1111/jofi.12185}
Maureen O'Hara, Chen Yao, and Mao Ye.
\newblock What's not there: Odd lots and market data.
\newblock \emph{The Journal of Finance}, 69\penalty0 (5):\penalty0 2199--2236,
  2014.
\newblock \doi{https://doi.org/10.1111/jofi.12185}.
\newblock URL \url{https://onlinelibrary.wiley.com/doi/abs/10.1111/jofi.12185}.

\bibitem[O'Shea and Nash(2015)]{oshea2015introduction}
Keiron O'Shea and Ryan Nash.
\newblock An introduction to convolutional neural networks, 2015.

\bibitem[Ribeiro et~al.(2016)Ribeiro, Singh, and Guestrin]{ribeiro2016why}
Marco~Tulio Ribeiro, Sameer Singh, and Carlos Guestrin.
\newblock "why should i trust you?" explaining the predictions of any
  classifier.
\newblock In \emph{Proceedings of the 22nd ACM SIGKDD International Conference
  on Knowledge Discovery and Data Mining}, pages 1135--1144, 2016.

\bibitem[Rossi(2018)]{rossi2018}
Alessio~S. Rossi.
\newblock Predicting stock market returns.
\newblock \emph{Journal of Financial Econometrics}, 16\penalty0 (4):\penalty0
  551--577, 2018.

\bibitem[Saar(2001)]{RePEc:oup:rfinst:v:14:y:2001:i:4:p:1153-81}
Gideon Saar.
\newblock Price impact asymmetry of block trades: An institutional trading
  explanation.
\newblock \emph{The Review of Financial Studies}, 14\penalty0 (4):\penalty0
  1153--81, 2001.
\newblock URL
  \url{https://EconPapers.repec.org/RePEc:oup:rfinst:v:14:y:2001:i:4:p:1153-81}.

\bibitem[Selvaraju et~al.(2019)Selvaraju, Cogswell, Das, Vedantam, Parikh, and
  Batra]{Selvaraju_2019}
Ramprasaath~R. Selvaraju, Michael Cogswell, Abhishek Das, Ramakrishna Vedantam,
  Devi Parikh, and Dhruv Batra.
\newblock Grad-cam: Visual explanations from deep networks via gradient-based
  localization.
\newblock \emph{International Journal of Computer Vision}, 128\penalty0
  (2):\penalty0 336–359, October 2019.
\newblock ISSN 1573-1405.
\newblock \doi{10.1007/s11263-019-01228-7}.
\newblock URL \url{http://dx.doi.org/10.1007/s11263-019-01228-7}.

\bibitem[Sirignano and Cont(2019)]{sirignano2019}
Justin Sirignano and R.~Cont.
\newblock Universal features of price formation in financial markets:
  Perspectives from deep learning.
\newblock \emph{Review of Financial Studies}, 32\penalty0 (5):\penalty0
  1459--1496, 2019.

\bibitem[Smilkov et~al.(2017)Smilkov, Thorat, Kim, Vi{\'e}gas, and
  Wattenberg]{smilkov2017smoothgrad}
D~Smilkov, N~Thorat, B~Kim, F~Vi{\'e}gas, and M~Wattenberg.
\newblock Smoothgrad: removing noise by adding noise; 2017. 10.48550.
\newblock \emph{arXiv preprint ARXIV.1706.03825}, 2017.

\bibitem[Srinivas and Fleuret(2019)]{srinivas2019fullgradient}
Suraj Srinivas and Fran{\c{c}}ois Fleuret.
\newblock Full-gradient representation for neural network visualization.
\newblock \emph{Advances in Neural Information Processing Iystems}, 32, 2019.

\bibitem[Srivastava et~al.(2014)Srivastava, Hinton, Krizhevsky, Sutskever, and
  Salakhutdinov]{JMLR:v15:srivastava14a}
Nitish Srivastava, Geoffrey Hinton, Alex Krizhevsky, Ilya Sutskever, and Ruslan
  Salakhutdinov.
\newblock Dropout: A simple way to prevent neural networks from overfitting.
\newblock \emph{Journal of Machine Learning Research}, 15\penalty0
  (56):\penalty0 1929--1958, 2014.
\newblock URL \url{http://jmlr.org/papers/v15/srivastava14a.html}.

\bibitem[Sundararajan et~al.(2017)Sundararajan, Taly, and
  Yan]{sundararajan2017axiomatic}
Mukund Sundararajan, Ankur Taly, and Qiqi Yan.
\newblock Axiomatic attribution for deep networks.
\newblock In \emph{International Conference on Machine Learning}, pages
  3319--3328. PMLR, 2017.

\bibitem[Sze et~al.(2017)Sze, Chen, Yang, and Emer]{sze2017efficient}
Vivienne Sze, Yu-Hsin Chen, Tien-Ju Yang, and Joel~S Emer.
\newblock Efficient processing of deep neural networks: A tutorial and survey.
\newblock \emph{Proceedings of the IEEE}, 105\penalty0 (12):\penalty0
  2295--2329, 2017.

\bibitem[Varian(2014)]{Varian2014Big}
Hal~R. Varian.
\newblock Big data: New tricks for econometrics.
\newblock \emph{Journal of Economic Perspectives}, 28\penalty0 (2):\penalty0
  3--28, 2014.

\bibitem[Watanabe(2023)]{watanabe2023treestructured}
Shuhei Watanabe.
\newblock Tree-structured parzen estimator: Understanding its algorithm
  components and their roles for better empirical performance.
\newblock \emph{arXiv preprint arXiv:2304.11127}, 2023.

\bibitem[Zeiler and Fergus(2014{\natexlab{a}})]{simonyan2014deep}
Matthew~D Zeiler and Rob Fergus.
\newblock Deep inside convolutional networks: Visualising image classification
  models and saliency maps.
\newblock In \emph{Computer Vision--ECCV 2014: 13th European Conference,
  Zurich, Switzerland, September 6-12, 2014, Proceedings, Part I 13}, pages
  818--833. Springer, 2014{\natexlab{a}}.

\bibitem[Zeiler and Fergus(2014{\natexlab{b}})]{zeiler2013visualizing}
Matthew~D Zeiler and Rob Fergus.
\newblock Visualizing and understanding convolutional networks.
\newblock In \emph{Computer Vision--ECCV 2014: 13th European Conference,
  Zurich, Switzerland, September 6-12, 2014, Proceedings, Part I 13}, pages
  818--833. Springer, 2014{\natexlab{b}}.

\bibitem[Zhang et~al.(2019)Zhang, He, Sra, and Jadbabaie]{zhang2020gradient}
Jingzhao Zhang, Tianxing He, Suvrit Sra, and Ali Jadbabaie.
\newblock Why gradient clipping accelerates training: A theoretical
  justification for adaptivity.
\newblock \emph{arXiv preprint arXiv:1905.11881}, 2019.

\bibitem[Zhou et~al.(2016)Zhou, Khosla, Lapedriza, Oliva, and
  Torralba]{zhou2015learning}
Bolei Zhou, Aditya Khosla, Agata Lapedriza, Aude Oliva, and Antonio Torralba.
\newblock Learning deep features for discriminative localization.
\newblock In \emph{Proceedings of the IEEE Conference on Computer Vision and
  Pattern Recognition}, pages 2921--2929, 2016.

\end{thebibliography}

\clearpage

\onehalfspacing

\section*{Tables} \label{sec:tab}
\addcontentsline{toc}{section}{Tables}
\centering

\begin{table}[ht!]
\centering
\begin{tabularx}{\linewidth}{>{\arraybackslash}X >{\centering\arraybackslash}X}
\toprule
Parameter & Value \\
\midrule
Number of Convolutional Layers & 4 \\
Number of Filters in Conv Layer 1 & 96 \\
Kernel Size in Conv Layer 1 & 4 \\
Initializer Method & he\_normal \\
Number of Filters in Conv Layer 2 & 70 \\
Kernel Size in Conv Layer 2 & 2 \\
Number of Filters in Conv Layer 3 & 35 \\
Kernel Size in Conv Layer 3 & 3 \\
Number of Dense Layers & 3 \\
Units in Dense Layer 1 & 304 \\
Dropout in Dense Layer 1 & 0.2729 \\
Units in Dense Layer 2 & 224 \\
Dropout in Dense Layer 2 & 0.3348 \\
Units in Dense Layer 3 & 544 \\
Dropout in Dense Layer 3 & 0.1340 \\
Learning Rate & 0.0010427 \\
Gradient Clipping Value & 1.0306 \\
Training Accuracy & 0.5774 \\
Validation Accuracy & 0.5657 \\
\bottomrule
\end{tabularx}
\caption{Best Neural Network Parameters and Performance}
\label{table:nn_parameters_performance}
\end{table}

\begin{table}[ht!]
\centering
{\small
\begin{tabularx}{\linewidth}{>{\arraybackslash}X >{\arraybackslash}X}
\toprule
Variable & Description \\
\midrule
\(w_{it} \times 1000\) & Response Variable - Influence score of trade \(i\) in window \(t\) \\ \\
Trade Size Indicators & Odd\_lot, 101\_to\_1000, 1000\_to\_10000, greater\_than\_10000 \\ \\
Time of Trade & Normalized time of trade within market hours \\ \\
Log Since Last Trade & Logarithm of time since last trade in nanoseconds \\ \\
Year Dummies & Indicators for the years 2018, 2019, 2020, 2021 \\ \\
Exchange Dummies & Indicators for each exchange venue \\ \\
Interaction Terms & Interaction terms between size, time, year, and exchange indicators \\ \\
Trade Price & Price at which the trade was executed \\ 
\bottomrule
\end{tabularx}
}
\caption{Description of Regression Variables}
\label{table:2}
\end{table}

\begin{table}[ht!]
\caption{Summary Statistics of Influence Scores by Trade Attribute}
\label{table:metrics}
\footnotesize
\begin{adjustbox}{angle=0}
\begin{tabularx}{\linewidth}{l *{6}{X}}
\toprule
Attribute & Count (million) & Max & Min & Mean & Std & Median \\
\midrule
Overall & 28.9 & 1601.27 & 0.00 & 52.90 & 70.70 & 27.69 \\
Predicted Negative & 9.33 & 1601.27 & 0.00 & 52.81 & 65.69 & 30.89 \\
Predicted Positive & 19.57 & 1550.34 & 0.00 & 52.95 & 72.98 & 26.31 \\
Stock & 25.59 & 1601.27 & 0.00 & 54.59 & 71.44 & 29.33 \\
ETF & 3.31 & 1550.34 & 0.00 & 39.89 & 63.29 & 16.79 \\ \\
odd\_lot & 11.76 & 1329.75 & 0.00 & 46.12 & 64.99 & 22.63 \\
round\_lot & 12.05 & 1601.27 & 0.00 & 53.23 & 70.11 & 28.34 \\
101\_to\_1000 & 4.66 & 1329.75 & 0.00 & 46.13 & 64.99 & 22.64 \\
1000\_to\_10000 & 0.41 & 1045.38 & 0.00 & 34.57 & 52.79 & 15.01 \\
greater\_than\_10000 & 0.02 & 721.73 & 0.00 & 29.98 & 45.75 & 12.66 \\ \\
2017 & 4.80 & 1262.92 & 0.00 & 48.97 & 65.87 & 25.65 \\
2018 & 6.76 & 1601.27 & 0.00 & 51.64 & 69.90 & 26.62 \\
2019 & 5.60 & 1539.12 & 0.00 & 54.08 & 71.21 & 28.57 \\
2020 & 0.34 & 1256.30 & 0.00 & 51.86 & 69.43 & 27.27 \\
2021 & 11.40 & 1550.34 & 0.00 & 54.77 & 72.82 & 28.87 \\ \\
NYSE & 2.09 & 1371.23 & 0.00 & 59.68 & 76.71 & 32.22 \\
NYSE American & 0.11 & 982.27 & 0.00 & 55.77 & 73.01 & 29.49 \\
NASDAQ OMX BX & 0.94 & 1101.30 & 0.00 & 43.15 & 55.64 & 23.82 \\
NYSE National & 0.37 & 1329.75 & 0.00 & 49.64 & 63.73 & 27.26 \\
ADF & 6.97 & 1539.12 & 0.00 & 43.75 & 59.91 & 22.51 \\
MIAX & 0.07 & 1099.10 & 0.00 & 77.76 & 99.15 & 41.31 \\
Cboe EDGA & 0.86 & 1212.29 & 0.00 & 45.12 & 58.31 & 24.58 \\
Cboe EDGX & 2.19 & 1550.34 & 0.00 & 55.87 & 73.77 & 30.12 \\
NYSE Chicago & 0.01 & 1017.73 & 0.00 & 51.78 & 73.33 & 24.51 \\
NYSE Arca & 2.79 & 1601.27 & 0.00 & 57.22 & 76.35 & 29.75 \\
CTS & 3.39 & 1398.10 & 0.00 & 60.13 & 79.45 & 31.70 \\
NASDAQ (C) & 3.25 & 1302.78 & 0.00 & 58.21 & 75.87 & 30.67 \\
MEMX & 0.37 & 1259.29 & 0.00 & 63.87 & 80.63 & 35.44 \\
IEX & 1.04 & 1498.15 & 0.00 & 53.84 & 69.57 & 29.01 \\
NASDAQ OMX PSX & 0.29 & 1050.79 & 0.00 & 54.71 & 71.21 & 29.19 \\
Cboe BYX & 1.45 & 1141.39 & 0.00 & 42.99 & 54.97 & 23.96 \\
Cboe BZX & 2.72 & 1487.88 & 0.00 & 57.81 & 75.48 & 30.60 \\
\bottomrule
\end{tabularx}
\end{adjustbox}
\end{table}

\begin{table}
\caption{Correlation Matrix of Trade Attributes}
\label{table:metrics}
{\footnotesize
\centering
\begin{adjustbox}{width= 17cm}
\begin{tabular}{llrrrrrrrrrrrrrrrrrrr}
\toprule
& Variable & 1 & 2 & 3 & 4 &5&6&7&8&9&10&11&12 \\
 \midrule
1.& ETF & 1.00 & -0.11 & 0.01 & -0.00 & 0.12 & 0.07 & 0.01 & 0.03 & -0.02 & -0.00 & -0.04 & 0.00 \\
2& Odd lot & -0.11 & 1.00 & -0.71 & 0.02 & -0.37 & -0.10 & -0.02 & -0.16 & -0.11 & -0.04 & 0.03 & 0.24 \\
3&round lot & 0.01 & -0.71 & 1.00 & 0.00 & -0.34 & -0.10 & -0.02 & 0.12 & 0.12 & 0.05 & -0.01 & -0.22 \\
4&log since last trade & -0.00 & 0.02 & 0.00 & 1.00 & -0.03 & 0.00 & 0.01 & 0.01 & 0.04 & 0.04 & 0.01 & -0.08 \\ \\
5&101 to 1000 & 0.12 & -0.37 & -0.34 & -0.03 & 1.00 & -0.05 & -0.01 & 0.06 & 0.00 & -0.00 & -0.02 & -0.04 \\
6&1000 to 10000 & 0.07 & -0.10 & -0.10 & 0.00 & -0.05 & 1.00 & -0.00 & 0.02 & -0.01 & -0.01 & -0.01 & 0.01 \\
7&greater than 10000 & 0.01 & -0.02 & -0.02 & 0.01 & -0.01 & -0.00 & 1.00 & 0.01 & 0.00 & -0.00 & -0.00 & -0.00 \\ \\
8&2017 & 0.03 & -0.16 & 0.12 & 0.01 & 0.06 & 0.02 & 0.01 & 1.00 & -0.22 & -0.21 & -0.05 & -0.38 \\
9&2018 & -0.02 & -0.11 & 0.12 & 0.04 & 0.00 & -0.01 & 0.00 & -0.22 & 1.00 & -0.24 & -0.06 & -0.45 \\
10&2019 & -0.00 & -0.04 & 0.05 & 0.04 & -0.00 & -0.01 & -0.00 & -0.21 & -0.24 & 1.00 & -0.05 & -0.42 \\
11&2020 & -0.04 & 0.03 & -0.01 & 0.01 & -0.02 & -0.01 & -0.00 & -0.05 & -0.06 & -0.05 & 1.00 & -0.10 \\
12&2021 & 0.00 & 0.24 & -0.22 & -0.08 & -0.04 & 0.01 & -0.00 & -0.38 & -0.45 & -0.42 & -0.10 & 1.00 \\ \\
13&NYSE & -0.07 & -0.01 & -0.04 & 0.03 & 0.07 & 0.01 & -0.00 & -0.02 & -0.00 & 0.03 & 0.00 & -0.01 \\
14&NYSE American & 0.02 & -0.01 & 0.01 & -0.02 & -0.00 & -0.00 & -0.00 & -0.02 & -0.01 & 0.01 & -0.00 & 0.02 \\
15&NASDAQ OMX BX & -0.02 & -0.02 & 0.05 & 0.03 & -0.02 & -0.01 & -0.00 & 0.06 & 0.07 & 0.00 & -0.01 & -0.10 \\
16&NYSE National & 0.02 & -0.01 & 0.02 & 0.01 & -0.01 & -0.00 & -0.00 & -0.05 & -0.03 & 0.04 & 0.00 & 0.03 \\
17&ADF & -0.02 & 0.01 & -0.07 & 0.40 & 0.06 & 0.06 & 0.02 & -0.02 & -0.05 & -0.06 & -0.02 & 0.11 \\
18&MIAX & 0.01 & -0.00 & 0.01 & -0.02 & -0.01 & 0.00 & -0.00 & -0.02 & -0.02 & -0.02 & -0.00 & 0.05 \\
19&Cboe EDGA & 0.01 & 0.00 & 0.02 & 0.00 & -0.03 & -0.02 & -0.00 & -0.00 & -0.03 & 0.04 & 0.02 & -0.02 \\
20&Cboe EDGX & 0.00 & 0.01 & -0.01 & -0.12 & 0.00 & -0.00 & -0.00 & 0.00 & 0.00 & -0.04 & -0.01 & 0.03 \\
21&NYSE Chicago & 0.01 & -0.00 & -0.01 & 0.00 & 0.00 & 0.01 & 0.01 & -0.01 & 0.01 & -0.00 & -0.00 & -0.00 \\
22&NYSE Arca & 0.10 & -0.02 & 0.02 & -0.11 & 0.00 & -0.00 & -0.00 & 0.03 & 0.00 & -0.01 & 0.00 & -0.02 \\
23&CTS & -0.06 & 0.04 & -0.02 & -0.11 & -0.03 & -0.02 & -0.01 & 0.01 & 0.07 & -0.03 & 0.03 & -0.04 \\
24&NASDAQ (C)  & 0.02 & 0.04 & -0.01 & -0.17 & -0.04 & -0.02 & -0.01 & -0.03 & -0.02 & 0.05 & 0.00 & -0.00 \\
25&MEMX & -0.01 & -0.01 & 0.02 & -0.07 & -0.01 & -0.00 & -0.00 & -0.05 & -0.06 & -0.05 & -0.01 & 0.12 \\
25&IEX & -0.02 & -0.04 & 0.05 & 0.05 & -0.01 & -0.01 & -0.00 & -0.03 & -0.01 & 0.02 & 0.00 & 0.01 \\
26&NASDAQ OMX PSX & 0.04 & -0.02 & 0.02 & -0.04 & 0.00 & -0.00 & -0.00 & 0.03 & 0.01 & -0.01 & -0.01 & -0.02 \\
27&Cboe BYX & 0.03 & -0.03 & 0.04 & 0.00 & -0.01 & -0.01 & -0.00 & 0.07 & 0.03 & 0.05 & -0.01 & -0.11 \\
28&Cboe BZX & 0.02 & -0.01 & 0.03 & -0.12 & -0.03 & -0.02 & -0.01 & 0.01 & 0.02 & 0.01 & 0.00 & -0.03 \\ \\
\bottomrule
\end{tabular}
\end{adjustbox}
}
\end{table}

\begin{table}[ht!]
\caption{Regression Results - Basic Attributes and Year}
\label{tab:regression-results}
\centering
\begin{adjustbox}{width=\textwidth, scale=1.2} % Adjust the scale factor as needed
\sisetup{
  detect-all,
  table-align-text-post = false,
  input-open-uncertainty  = ,
  input-close-uncertainty = ,
  table-space-text-pre   = (,
  table-space-text-post  = {***},
  table-align-text-pre   = false,
  table-number-alignment = center, % Center values in the right three columns
  table-figures-integer  = 3,
  table-figures-decimal  = 2,
}
\begin{tabular}{l *{3}{S[table-format=-1.2(3), table-space-text-pre=(, table-space-text-post=***]}}
\toprule
\multirow{2}{*}{\textbf{Variable}} & \multicolumn{3}{c}{\textbf{Model}} \\
\cmidrule(lr){2-4}
& {All Observations} & {Trades from windows Predicted Positive} & {Trades from windows Predicted Negative} \\
\midrule
Constant              & 55.50*** &  55.32*** & 63.85*** \\
                      & (0.37)   &  (0.48) & (0.46)   \\
ETF                   & -13.14*** & -8.17*** & -29.94*** \\
                      & (0.37)    & (0.46) & (0.34) \\
Time of Trade         & 6.89***  & 2.49*** & 12.86*** \\
                      & (0.37)   & (0.49) & (0.45) \\
Log Since Last Trade  & -0.47*** & -0.45*** & -0.63*** \\
                      & (0.01)   & (0.01) & (0.01) \\
Predicted             & 2.36***  &  &  \\
                      & (0.22)   &  &  \\
2018                  & -1.76*** & -0.53 & -3.30*** \\
                      & (0.37)   & (0.50) & (0.44) \\
2019                  & 0.66*    & 1.66*** & -0.55 \\
                      & (0.36)   & (0.48) & (0.44) \\
2020                  & -4.82*** & -2.01 & -5.34*** \\
                      & (0.94)   & (1.30) & (0.98) \\
2021                  & 2.44***  & 6.34*** & -5.26*** \\
                      & (0.33)   & (0.45) & (0.40) \\
Trade Price           & 0.01***  & 0.02*** & -0.03*** \\
                      & (0.001)  & (0.001) & (0.001) \\
\midrule
R-squared             & {0.008} & {0.010} & {0.033} \\
Adjusted R-squared    & {0.008} & {0.010} & {0.033} \\
Number of Observations & {28,900,000} & {20,128,561} & {8,771,439} \\
\bottomrule
\end{tabular}
\end{adjustbox}
\footnotesize Note: Standard errors are clustered at the window level. Significance levels: \textsuperscript{*}p<0.05, \textsuperscript{**}p<0.01, \textsuperscript{***}p<0.001.
\end{table}

\begin{table}[ht!]
\caption{Regression Results - Trade Sizes Included}
\label{tab:regression-results}
\centering
\begin{adjustbox}{width=\textwidth, scale=1.2} % Adjust the scale factor as needed
\sisetup{
  detect-all,
  table-align-text-post = false,
  input-open-uncertainty  = ,
  input-close-uncertainty = ,
  table-space-text-pre   = (,
  table-space-text-post  = {***},
  table-align-text-pre   = false,
  table-number-alignment = center, % Center values in the right three columns
  table-figures-integer  = 3,
  table-figures-decimal  = 2,
}
\begin{tabular}{l *{3}{S[table-format=-1.2(3), table-space-text-pre=(, table-space-text-post=***]}}
\toprule
\multirow{2}{*}{\textbf{Variable}} & \multicolumn{3}{c}{\textbf{Model}} \\
\cmidrule(lr){2-4}
& {All Observations} & {Trades from windows Predicted Positive} & {Trades from windows Predicted Negative} \\
\midrule
Constant              & 56.84*** & 56.05***      & 65.29*** \\
                      & (0.372)  & (0.475)      & (0.459)  \\
ETF                   & -11.66*** & -6.46*** & -28.71*** \\
                      & (0.368)   & (0.457) & (0.339)  \\
Time of Trade         & 6.59*** & 2.10*** & 12.61*** \\
                      & (0.370)  & (0.487) & (0.450)  \\
Log Since Last Trade  & -0.48*** & -0.45*** & -0.64*** \\
                      & (0.008)  & (0.009) & (0.012)  \\
Predicted             & 1.95*** &  &  \\
                      & (0.216)  &  &  \\
2018                  & -1.90*** & -0.79 & -3.29*** \\
                      & (0.372)  & (0.499) & (0.442)  \\
2019                  & 0.35 & 1.05*** & -0.51 \\
                      & (0.362)  & (0.482) & (0.437)  \\
2020                  & -4.98*** & -2.71*** & -5.04*** \\
                      & (0.936)  & (1.285) & (0.978)  \\
2021                  & 1.66*** & 5.00*** & -5.33*** \\
                      & (0.330)  & (0.447) & (0.397)  \\
Trade Price           & 0.01*** & 0.02*** & -0.04*** \\
                      & (0.001)  & (0.001) & (0.001)  \\
Odd lot               & 1.89*** & 3.47*** & -0.12 \\
                      & (0.096)  & (0.132) & (0.126)  \\
101 to 1000           & -5.76*** & -6.27*** & -4.79*** \\
                      & (0.082)  & (0.108) & (0.114)  \\
1000 to 10000         & -16.88*** & -19.60*** & -12.04*** \\
                      & (0.194)  & (0.259) & (0.272)  \\
Greater than 10000    & -23.11*** & -26.20*** & -17.40*** \\
                      & (0.421)  & (0.541) & (0.689)  \\
\midrule
R-squared             & {0.010} & {0.012} & {0.034} \\
Adjusted R-squared    & {0.010} & {0.012} & {0.034} \\
Number of Observations & {28,900,000} & {20,128,561} & {8,771,439} \\
\bottomrule
\end{tabular}
\end{adjustbox}
\footnotesize Note: Standard errors are clustered at the window level. Significance levels: \textsuperscript{*}p<0.05, \textsuperscript{**}p<0.01, \textsuperscript{***}p<0.001.
\end{table}

\begin{table}[ht!]
\caption{Regression Results - Trade Size, Year Interaction Effects Included}
\label{tab:regression-results}
\centering
\begin{adjustbox}{width=\textwidth, scale=1.2} % Adjust the scale factor as needed
\sisetup{
  detect-all,
  table-align-text-post = false,
  input-open-uncertainty  = ,
  input-close-uncertainty = ,
  table-space-text-pre   = (,
  table-space-text-post  = {***},
  table-align-text-pre   = false,
  table-number-alignment = center, % Center values in the right three columns
  table-figures-integer  = 3,
  table-figures-decimal  = 2,
}
\begin{tabular}{l *{3}{S[table-format=-1.2(3), table-space-text-pre=(, table-space-text-post=***]}}
\toprule
\multirow{2}{*}{\textbf{Variable}} & \multicolumn{3}{c}{\textbf{Model}} \\
\cmidrule(lr){2-4}
& {All Observations} & {Trades from windows Predicted Positive} & {Trades from windows Predicted Negative} \\
\midrule
Intercept                          & 56.93***  & 56.42***       & 64.98***  \\
                                  & (0.38)    &  (0.48)      & (0.47)     \\
ETF                                & -11.71*** & -6.52*** & -28.84*** \\
                                  & (0.37)    & (0.46) & (0.34)     \\
Time\_of\_Trade                    & 6.59***   & 2.09***  & 12.55***  \\
                                  & (0.37)    & (0.49) & (0.45)     \\
log\_since\_last\_trade\_ns        & -0.48***  & -0.45*** & -0.64***  \\
                                  & (0.01)    & (0.01) & (0.01)     \\
predicted                          & 1.97***   &  &  \\
                                  & (0.22)    &  &  \\
2018                              & -2.10***  & -1.28**  & -2.76***  \\
                                  & (0.38)    & (0.51) & (0.46)     \\
2019                              & 0.66      & 1.63***  & -0.66     \\
                                  & (0.37)    & (0.50) & (0.46)     \\
2020                              & -7.63***  & -5.72*** & -5.91***  \\
                                  & (0.93)    & (1.31) & (1.06)     \\
2021                              & 1.47***   & 4.05***  & -4.43***  \\
                                  & (0.34)    & (0.46) & (0.42)     \\
Trade\_Price                      & 0.01***   & 0.02***  & -0.04***  \\
                                  & (0.00)    & (0.00) & (0.00)     \\
Odd\_lot                          & 1.44***   & 1.86***  & 1.44***   \\
                                  & (0.23)    & (0.30) & (0.28)     \\
101\_to\_1000                     & -5.51***  & -5.64*** & -4.89***  \\
                                  & (0.18)    & (0.24) & (0.24)     \\
1000\_to\_10000                   & -16.73*** & -17.93*** & -12.52*** \\
                                  & (0.42)    & (0.56) & (0.56)     \\
greater\_than\_10000              & -22.76*** & -24.22*** & -17.06*** \\
                                  & (0.94)    & (1.18) & (1.53)     \\
interaction\_Odd\_lot\_2018       & -0.46     & 1.03**   & -3.40***  \\
                                  & (0.34)    & (0.43) & (0.44)     \\
interaction\_Odd\_lot\_2019       & -1.28***  & -1.27*** & -0.72     \\
                                  & (0.31)    & (0.40) & (0.40)     \\
interaction\_Odd\_lot\_2020       & 6.17***   & 7.95***  & -3.76***  \\
                                  & (0.92)    & (1.13) & (0.98)     \\
interaction\_Odd\_lot\_2021       & 1.30***   & 3.33***  & -2.07***  \\
                                  & (0.28)    & (0.36) & (0.35)     \\
interaction\_101\_to\_1000\_2018  & 1.72***   & 1.23***  & 1.44***   \\
                                  & (0.26)    & (0.34) & (0.35)     \\
interaction\_101\_to\_1000\_2019  & 0.94***   & 0.61*    & 0.97***   \\
                                  & (0.25)    & (0.33) & (0.33)     \\
interaction\_101\_to\_1000\_2020  & -0.58     & -3.60*** & 5.19***   \\
                                  & (0.64)    & (0.85) & (0.79)     \\
interaction\_101\_to\_1000\_2021  & -2.52***  & -2.97*** & -1.75***  \\
                                  & (0.23)    & (0.31) & (0.32)     \\
interaction\_1000\_to\_10000\_2018& 4.26***   & 3.66***  & 2.03**    \\
                                  & (0.56)    & (0.73) & (0.82)     \\
interaction\_1000\_to\_10000\_2019& 3.33***   & 2.64***  & 2.29***   \\
                                  & (0.54)    & (0.71) & (0.81)     \\
interaction\_1000\_to\_10000\_2020& 5.32***   & -0.78    & 11.77***  \\
                                  & (1.16)    & (1.46) & (1.52)     \\
interaction\_1000\_to\_10000\_2021& -5.22***  & -7.66*** & -2.81***  \\
                                  & (0.52)    & (0.69) & (0.75)     \\
interaction\_greater\_than\_10000\_2018 & 6.23*** & 7.39*** & -0.59     \\
                                  & (1.34)    & (1.67) & (2.23)     \\
interaction\_greater\_than\_10000\_2019 & 5.26*** & 5.51*** & 2.66      \\
                                  & (1.25)    & (1.55) & (2.13)     \\
interaction\_greater\_than\_10000\_2020 & 13.00*** & 9.52*** & 17.13*** \\
                                  & (2.73)    & (3.81) & (3.90)     \\
interaction\_greater\_than\_10000\_2021 & -6.57*** & -9.63*** & -3.34*    \\
                                  & (1.11)    & (1.37) & (1.91)     \\
\midrule
Observations                      & {28900000} & {20128561} & {8771439} \\
R-squared                         & {0.010}    & {0.013}    & {0.035}   \\
\bottomrule
\end{tabular}
\end{adjustbox}
\footnotesize Note: Standard errors are clustered at the window level. Significance levels: \textsuperscript{*}p<0.05, \textsuperscript{**}p<0.01, \textsuperscript{***}p<0.001.
\end{table}

\begin{table}[ht!]
\caption{Regression Results - Volume, Year Interaction Measures Included }
\label{tab:regression-results-10-12}
\centering
\begin{adjustbox}{width=\textwidth, scale=1.2} % Adjust the scale factor as needed
\sisetup{
  detect-all,
  table-align-text-post = false,
  input-open-uncertainty  = ,
  input-close-uncertainty = ,
  table-space-text-pre   = (,
  table-space-text-post  = {***},
  table-align-text-pre   = false,
  table-number-alignment = center, % Center values in the right three columns
  table-figures-integer  = 3,
  table-figures-decimal  = 2,
}
\begin{tabular}{l *{3}{S[table-format=-1.2(3), table-space-text-pre=(, table-space-text-post=***]}}
\toprule
\multirow{2}{*}{\textbf{Variable}} & \multicolumn{3}{c}{\textbf{Model}} \\
\cmidrule(lr){2-4}
& {All Observations} & {Trades from windows Predicted Positive} & {Trades from windows Predicted Negative} \\
\midrule
Intercept                          & 56.51***  &  54.92***        & 65.13***  \\
                                  & (0.46)    &  (0.59)         & (0.63)    \\
ETF                                & -11.69*** & -6.48*** & -28.82*** \\
                                  & (0.37)    & (0.46)   & (0.34)    \\
Time\_of\_Trade                    & 6.58***   & 2.08***  & 12.55***  \\
                                  & (0.37)    & (0.49)   & (0.45)    \\
log\_since\_last\_trade\_ns        & -0.45***  & -0.35*** & -0.65***  \\
                                  & (0.02)    & (0.03)   & (0.03)    \\
predicted                          & 1.97***   &  &  \\
                                  & (0.22)    &    &  \\
2018                              & -2.95***  & -1.70**  & -3.83***  \\
                                  & (0.55)    & (0.71)   & (0.80)    \\
2019                              & -0.25     & 0.54     & -1.41     \\
                                  & (0.54)    & (0.69)   & (0.80)    \\
2020                              & -10.99*** & -7.77*** & -10.69*** \\
                                  & (1.22)    & (1.68)   & (1.67)    \\
2021                              & 3.04***   & 7.69***  & -3.95***  \\
                                  & (0.48)    & (0.63)   & (0.69)    \\
Trade\_Price                      & 0.01***   & 0.02***  & -0.04***  \\
                                  & (0.00)    & (0.00)   & (0.00)    \\
Odd\_lot                          & 1.45***   & 1.85***  & 1.44***   \\
                                  & (0.23)    & (0.30)   & (0.28)    \\
101\_to\_1000                     & -5.50***  & -5.59*** & -4.90***  \\
                                  & (0.18)    & (0.24)   & (0.24)    \\
1000\_to\_10000                   & -16.73*** & -17.93*** & -12.53*** \\
                                  & (0.42)    & (0.56)   & (0.56)    \\
greater\_than\_10000              & -22.79*** & -24.34*** & -17.07*** \\
                                  & (0.94)    & (1.17)   & (1.53)    \\
interaction\_Odd\_lot\_2018       & -0.47     & 1.01**   & -3.39***  \\
                                  & (0.34)    & (0.43)   & (0.44)    \\
interaction\_Odd\_lot\_2019       & -1.25***  & -1.21*** & -0.71     \\
                                  & (0.31)    & (0.40)   & (0.40)    \\
interaction\_Odd\_lot\_2020       & 6.30***   & 8.12***  & -3.69***  \\
                                  & (0.92)    & (1.13)   & (0.98)    \\
interaction\_Odd\_lot\_2021       & 1.34***   & 3.40***  & -2.05***  \\
                                  & (0.28)    & (0.36)   & (0.35)    \\
interaction\_101\_to\_1000\_2018  & 1.74***   & 1.23***  & 1.49***   \\
                                  & (0.25)    & (0.34)   & (0.34)    \\
interaction\_101\_to\_1000\_2019  & 0.98***   & 0.66*    & 1.00***   \\
                                  & (0.25)    & (0.33)   & (0.33)    \\
interaction\_101\_to\_1000\_2020  & -0.41     & -3.52*** & 5.45***   \\
                                  & (0.64)    & (0.85)   & (0.79)    \\
interaction\_101\_to\_1000\_2021  & -2.53***  & -2.99*** & -1.74***  \\
                                  & (0.23)    & (0.31)   & (0.32)    \\
interaction\_1000\_to\_10000\_2018& 4.23***   & 3.59***  & 2.02**    \\
                                  & (0.56)    & (0.73)   & (0.82)    \\
interaction\_1000\_to\_10000\_2019& 3.36***   & 2.69***  & 2.32***   \\
                                  & (0.54)    & (0.71)   & (0.81)    \\
interaction\_1000\_to\_10000\_2020& 5.67***   & -0.49    & 12.17***  \\
                                  & (1.15)    & (1.45)   & (1.53)    \\
interaction\_1000\_to\_10000\_2021& -5.12***  & -7.46*** & -2.75***  \\
                                  & (0.52)    & (0.69)   & (0.75)    \\
interaction\_greater\_than\_10000\_2018 & 6.04*** & 7.15*** & -0.72     \\
                                  & (1.34)    & (1.67)   & (2.23)    \\
interaction\_greater\_than\_10000\_2019 & 5.13*** & 5.24*** & 2.61      \\
                                  & (1.25)    & (1.55)   & (2.13)    \\
interaction\_greater\_than\_10000\_2020 & 12.85*** & 9.33**  & 17.14*** \\
                                  & (2.72)    & (3.77)   & (3.95)    \\
interaction\_greater\_than\_10000\_2021 & -6.34*** & -9.16*** & -3.22*    \\
                                  & (1.11)    & (1.37)   & (1.91)    \\
interaction\_2018\_time\_diff     & 0.06*     & 0.03     & 0.08      \\
                                  & (0.03)    & (0.03)   & (0.04)    \\
interaction\_2019\_time\_diff     & 0.06**    & 0.07**   & 0.05      \\
                                  & (0.03)    & (0.03)   & (0.04)    \\
interaction\_2020\_time\_diff     & 0.23***   & 0.13*    & 0.34***   \\
                                  & (0.05)    & (0.07)   & (0.08)    \\
interaction\_2021\_time\_diff     & -0.11***  & -0.26*** & -0.04     \\
                                  & (0.02)    & (0.03)   & (0.04)    \\
\midrule
Observations                      & {28900000} & {20128561} & {8771439} \\
R-squared                         & {0.010}    & {0.013}    & {0.035}   \\
\bottomrule
\end{tabular}
\end{adjustbox}
\footnotesize Note: Standard errors are clustered at the window level. Significance levels: \textsuperscript{*}p<0.05, \textsuperscript{**}p<0.01, \textsuperscript{***}p<0.001.
\end{table}

\begin{table}[ht!]
\caption{Regression Results: Full Model with Venue Effects Included}
\label{tab:regression-results-13-15}
\centering
\begin{adjustbox}{width=\textwidth, scale=1.2} % Adjust the scale factor as needed
\sisetup{
  detect-all,
  table-align-text-post = false,
  input-open-uncertainty  = ,
  input-close-uncertainty = ,
  table-space-text-pre   = (,
  table-space-text-post  = {***},
  table-align-text-pre   = false,
  table-number-alignment = center, % Center values in the right three columns
  table-figures-integer  = 3,
  table-figures-decimal  = 2,
}
\begin{tabular}{l *{3}{S[table-format=-1.2(3), table-space-text-pre=(, table-space-text-post=***]}}
\toprule
\multirow{2}{*}{\textbf{Variable}} & \multicolumn{3}{c}{\textbf{Model}} \\
\cmidrule(lr){2-4}
& {All Observations} & {Trades from windows Predicted Positive} & {Trades from windows Predicted Negative} \\
\midrule
Intercept                          & 56.84***  & 53.66***         & 68.37***  \\
                                  & (0.48)    & (0.60)         & (0.65)    \\
ETF                                & -11.76*** & -6.47*** & -29.14*** \\
                                  & (0.37)    & (0.45)   & (0.34)    \\
Time\_of\_Trade                    & 6.90***   & 2.37***  & 12.76***  \\
                                  & (0.37)    & (0.49)   & (0.45)    \\
log\_since\_last\_trade\_ns        & 0.03      & 0.10***  & -0.09***  \\
                                  & (0.02)    & (0.03)   & (0.03)    \\
predicted                          & 1.45***   &  &  \\
                                  & (0.22)    &    &  \\
2018                              & -3.42***  & -2.03**  & -4.50***  \\
                                  & (0.55)    & (0.71)   & (0.80)    \\
2019                              & 0.55      & 1.62*    & -1.12     \\
                                  & (0.54)    & (0.69)   & (0.80)    \\
2020                              & -10.84*** & -7.65*** & -11.31*** \\
                                  & (1.22)    & (1.67)   & (1.70)    \\
2021                              & 1.52***   & 6.82***  & -6.99***  \\
                                  & (0.48)    & (0.63)   & (0.69)    \\
Trade\_Price                      & 0.01***   & 0.02***  & -0.03***  \\
                                  & (0.00)    & (0.00)   & (0.00)    \\
Odd\_lot                          & 0.58*     & 1.06***  & 0.37      \\
                                  & (0.23)    & (0.30)   & (0.28)    \\
101\_to\_1000                     & -5.14***  & -5.14*** & -4.81***  \\
                                  & (0.18)    & (0.24)   & (0.24)    \\
1000\_to\_10000                   & -14.95*** & -15.83*** & -11.43*** \\
                                  & (0.41)    & (0.55)   & (0.56)    \\
greater\_than\_10000              & -20.28*** & -21.47*** & -15.15*** \\
                                  & (0.95)    & (1.20)   & (1.53)    \\
interaction\_Odd\_lot\_2018       & -0.37     & 1.09**   & -3.17***  \\
                                  & (0.34)    & (0.43)   & (0.44)    \\
interaction\_Odd\_lot\_2019       & -1.26***  & -1.22*** & -0.59     \\
                                  & (0.31)    & (0.40)   & (0.40)    \\
interaction\_Odd\_lot\_2020       & 5.51***   & 7.03***  & -3.96***  \\
                                  & (0.92)    & (1.12)   & (0.98)    \\
interaction\_Odd\_lot\_2021       & 2.35***   & 4.15***  & -0.44     \\
                                  & (0.28)    & (0.36)   & (0.35)    \\
interaction\_101\_to\_1000\_2018  & 1.87***   & 1.34***  & 1.58***   \\
                                  & (0.25)    & (0.33)   & (0.34)    \\
interaction\_101\_to\_1000\_2019  & 0.81***   & 0.48     & 0.79**    \\
                                  & (0.25)    & (0.33)   & (0.33)    \\
interaction\_101\_to\_1000\_2020  & -0.29     & -3.10*** & 5.10***   \\
                                  & (0.62)    & (0.83)   & (0.77)    \\
interaction\_101\_to\_1000\_2021  & -1.81***  & -2.48*** & -0.49     \\
                                  & (0.23)    & (0.30)   & (0.31)    \\
interaction\_1000\_to\_10000\_2018& 4.87***   & 3.99***  & 2.93***   \\
                                  & (0.55)    & (0.72)   & (0.81)    \\
interaction\_1000\_to\_10000\_2019& 2.86***   & 2.18**   & 1.66*     \\
                                  & (0.54)    & (0.70)   & (0.82)    \\
interaction\_1000\_to\_10000\_2020& 4.00***   & -1.45    & 9.75***   \\
                                  & (1.13)    & (1.45)   & (1.47)    \\
interaction\_1000\_to\_10000\_2021& -3.37***  & -6.13*** & -0.04     \\
                                  & (0.51)    & (0.67)   & (0.74)    \\
interaction\_greater\_than\_10000\_2018 & 8.08*** & 8.80*** & 1.67     \\
                                  & (1.36)    & (1.70)   & (2.24)    \\
interaction\_greater\_than\_10000\_2019 & 5.68*** & 5.82*** & 2.77     \\
                                  & (1.26)    & (1.58)   & (2.13)    \\
interaction\_greater\_than\_10000\_2020 & 12.50*** & 9.37**  & 15.52*** \\
                                  & (2.78)    & (3.90)   & (3.98)    \\
interaction\_greater\_than\_10000\_2021 & -4.57*** & -8.17*** & 0.25     \\
                                  & (1.13)    & (1.41)   & (1.88)    \\
interaction\_2018\_time\_diff     & 0.07***   & 0.03     & 0.10**    \\
                                  & (0.03)    & (0.03)   & (0.04)    \\
interaction\_2019\_time\_diff     & -0.00     & -0.01    & 0.01      \\
                                  & (0.03)    & (0.03)   & (0.04)    \\
interaction\_2020\_time\_diff     & 0.20***   & 0.12*    & 0.34***   \\
                                  & (0.05)    & (0.07)   & (0.08)    \\
interaction\_2021\_time\_diff     & -0.05**   & -0.22*** & 0.09**    \\
                                  & (0.02)    & (0.03)   & (0.04)    \\
\bottomrule
\end{tabular}
\end{adjustbox}
\footnotesize Note: Standard errors are clustered at the window level. Significance levels: \textsuperscript{*}p<0.05, \textsuperscript{**}p<0.01, \textsuperscript{***}p<0.001.
\end{table}

\begin{table}[ht!]
\caption{Regression Results: Full Model with Venue Effects Included}
\centering
\begin{adjustbox}{width=\textwidth, scale=1.2} % Adjust the scale factor as needed
\sisetup{
  detect-all,
  table-align-text-post = false,
  input-open-uncertainty  = ,
  input-close-uncertainty = ,
  table-space-text-pre   = (,
  table-space-text-post  = {***},
  table-align-text-pre   = false,
  table-number-alignment = center, % Center values in the right three columns
  table-figures-integer  = 3,
  table-figures-decimal  = 2,
}
\begin{tabular}{l *{3}{S[table-format=-1.2(3), table-space-text-pre=(, table-space-text-post=***]}}
\toprule
\multirow{2}{*}{\textbf{Variable}} & \multicolumn{3}{c}{\textbf{Model}} \\
\cmidrule(lr){2-4}
& {All Observations} & {Trades from windows Predicted Positive} & {Trades from windows Predicted Negative} \\
\midrule
                                  
NYSE American                    & -5.95***  & -5.23*** & -9.74***  \\
                                  & (0.41)    & (0.51)   & (0.68)    \\
NASDAQ OMX BX                    & -15.76*** & -13.60***& -21.76*** \\
                                  & (0.18)    & (0.23)   & (0.27)    \\
NYSE National                    & -10.99*** & -10.37***& -14.48*** \\
                                  & (0.25)    & (0.32)   & (0.37)    \\
ADF                               & -15.73*** & -13.49***& -21.27*** \\
                                  & (0.17)    & (0.22)   & (0.24)    \\
MIAX                              & 11.31***  & 10.31*** & 12.44***  \\
                                  & (0.59)    & (0.74)   & (0.91)    \\
Cboe EDGA                        & -14.62*** & -12.93***& -19.70*** \\
                                  & (0.18)    & (0.23)   & (0.27)    \\
Cboe EDGX                        & -1.84***  & 0.18     & -5.57***  \\
                                  & (0.19)    & (0.23)   & (0.27)    \\
NYSE Chicago                      & -5.82***  & -2.63*   & -13.00*** \\
                                  & (1.04)    & (1.33)   & (1.34)    \\
NYSE Arca                        & -1.07***  & 0.01     & -3.74***  \\
                                  & (0.17)    & (0.22)   & (0.25)    \\
CTS                               & -1.79***  & 1.73***  & -9.37***  \\
                                  & (0.28)    & (0.37)   & (0.33)    \\
NASDAQ (C)                       & -0.21     & -0.60*** & 0.95***   \\
                                  & (0.13)    & (0.16)   & (0.22)    \\
MEMX                             & -0.46     & -3.20*** & 2.63***   \\
                                  & (0.33)    & (0.42)   & (0.49)    \\
IEX                              & -6.90***  & -5.87*** & -10.15*** \\
                                  & (0.19)    & (0.23)   & (0.28)    \\
NASDAQ OMX PSX                   & -4.54***  & -4.31*** & -6.39***  \\
                                  & (0.25)    & (0.32)   & (0.37)    \\
Cboe BYX                         & -15.36*** & -13.77***& -20.44*** \\
                                  & (0.18)    & (0.23)   & (0.26)    \\
Cboe BZX                         & -1.74***  & -1.09*** & -4.07***  \\
                                  & (0.20)   & (0.24)   & (0.24)    \\
\bottomrule
\end{tabular}
\end{adjustbox}
\vspace{0in}
\footnotesize Note: Standard errors are clustered at the window level. Significance levels: \textsuperscript{*}p<0.05, \textsuperscript{**}p<0.01, \textsuperscript{***}p<0.001.
\end{table}

\clearpage

\section*{Figures} \label{sec:fig}
\addcontentsline{toc}{section}{Figures}

\begin{figure}[ht!]
    \centering
    \includegraphics[width=10cm, height=18cm]{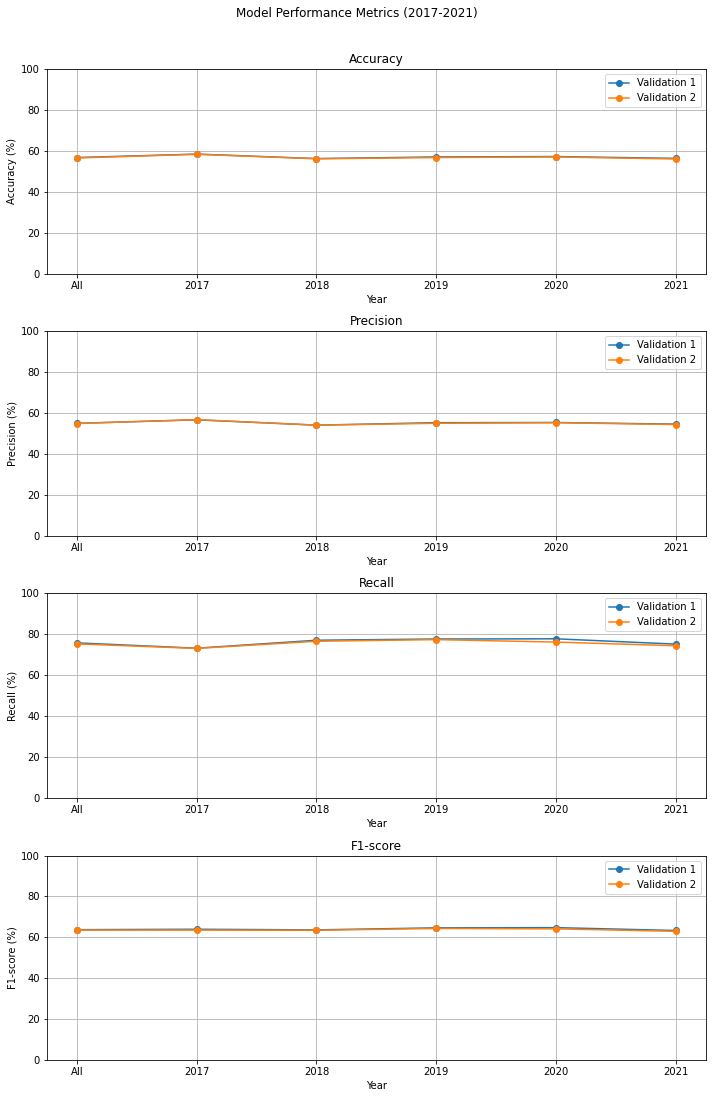}
    \caption{Prediction accuracies generated from applying optimized predictor to two separate valiation sets, by year}
    \label{fig:fig1}
\end{figure}

\begin{figure}[ht!]
    \centering
    \begin{adjustbox}{width=\textwidth}
        \includegraphics[width=10cm, height=8cm, angle = 270]{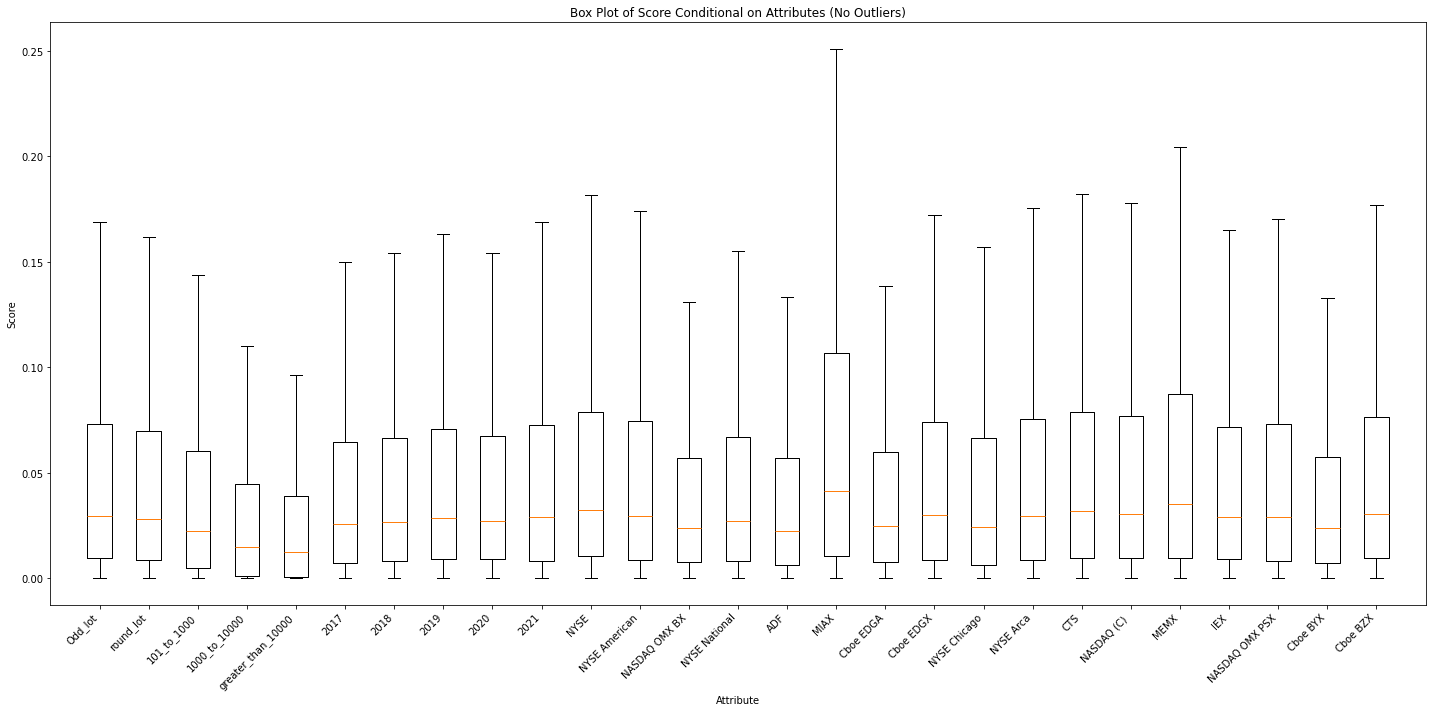}
    \end{adjustbox}
    \caption{Visualizing conditional distributions of influence scores by trade attribute category}
    \label{fig:fig1}
\end{figure}

\begin{figure}[ht!]
    \centering
    \includegraphics[width=10cm, height=8cm]{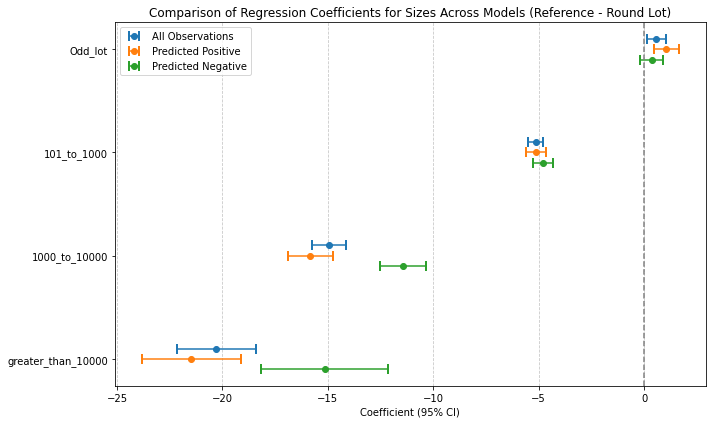}
    \caption{A Plot for regression coefficients and confidence intervals for basic trade attributes}
    \label{fig:fig1}
\end{figure}

\begin{figure}[ht!]
    \centering
    \includegraphics[width=10cm, height=8cm]{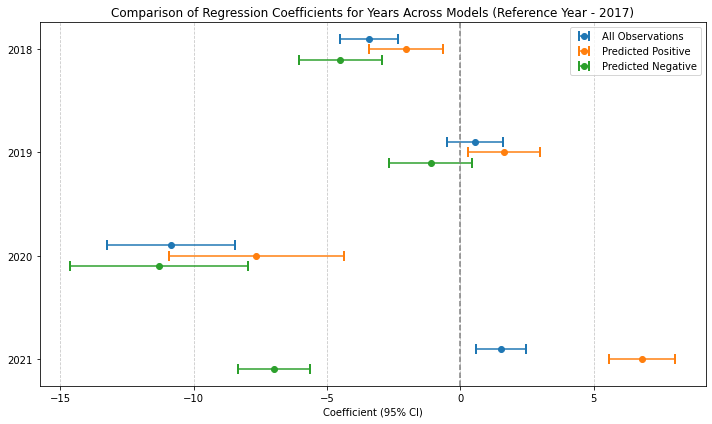}
    \caption{A Plot for regression coefficients and confidence intervals for year of trade}
    \label{fig:fig1}
\end{figure}

\begin{figure}[ht!]
    \centering
    \includegraphics[width=15cm, height=15cm]{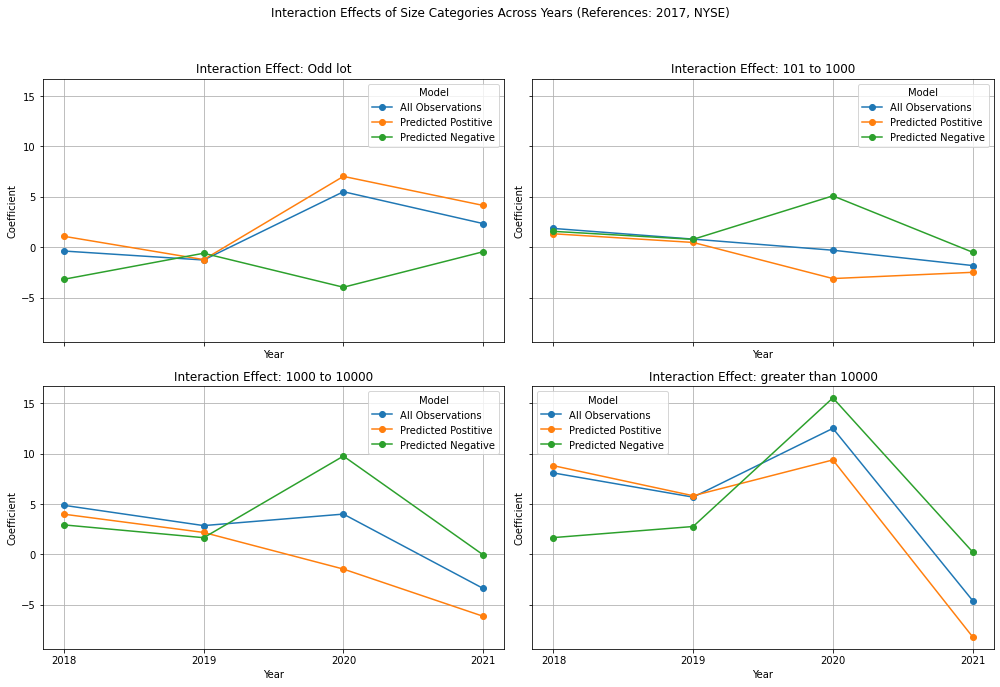}
    \caption{A Plot for regression coefficients and confidence intervals for year of trade, trade size interaction effects}
    \label{fig:fig1}
\end{figure}

\begin{figure}[ht!]
    \centering
    \includegraphics[width=10cm, height=8cm]{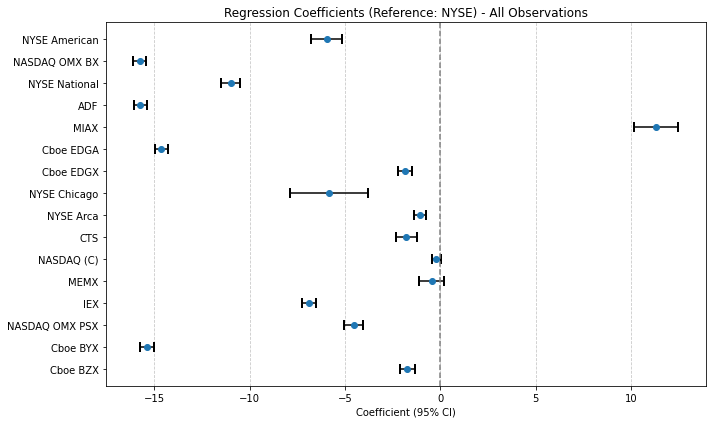}
    \caption{A Plot for regression coefficients and confidence intervals for trade venues for all trade windows}
    \label{fig:fig1}
\end{figure}

\begin{figure}[ht!]
    \centering
    \includegraphics[width=10cm, height=8cm]{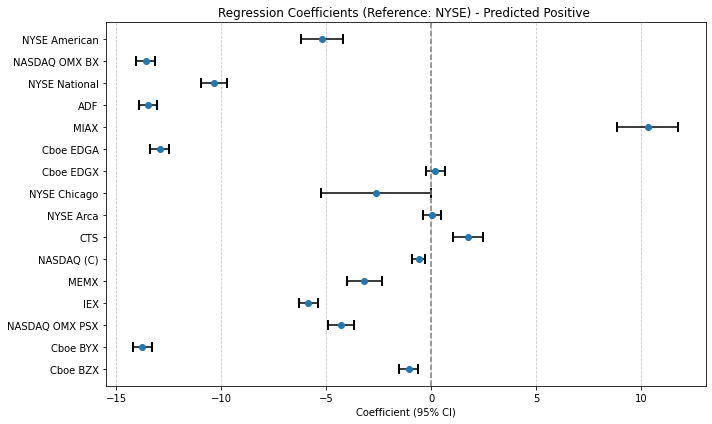}
    \caption{A Plot for regression coefficients and confidence intervals for trade venues for trade windows with positive predicted future price movements}
    \label{fig:fig1}
\end{figure}

\begin{figure}[ht!]
    \centering
    \includegraphics[width=10cm, height=8cm]{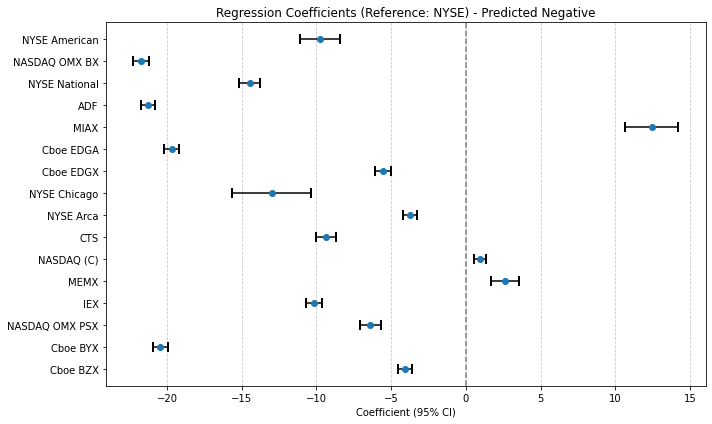}
    \caption{A Plot for regression coefficients and confidence intervals for trade venues for trade windows with negative predicted future price movements}
    \label{fig:fig1}
\end{figure}
%\begin{figure}[hp]
%  \centering
%  \includegraphics[width=.6\textwidth]{../fig/placeholder.pdf}
%  \caption{Placeholder}
%  \label{fig:placeholder}
%\end{figure}

\clearpage

\end{document}